\documentclass[ journal]{IEEEtran}

\IEEEoverridecommandlockouts
\usepackage[T1]{fontenc}
\usepackage{amsmath}
\usepackage{amsthm}
\usepackage[cmintegrals]{newtxmath}
\usepackage{bm}
\usepackage{bbm}
\usepackage{comment}
\usepackage{graphicx}
\usepackage{calc}
\usepackage{epstopdf}
\newtheorem{theorem}{Theorem}
\newtheorem{lemma}{Lemma}
\newtheorem{remark}{Remark}

\newtheorem{corollary}{Corollary}

\newtheorem{assumption}{Assumption}
\usepackage{verbatim}
\usepackage{color}
\usepackage{array}
\usepackage{cite}
\usepackage{stfloats}	

\usepackage{soul}

\usepackage{xcolor}
\usepackage{floatrow}
\usepackage{lipsum}
\usepackage[ruled,norelsize]{algorithm2e}
\usepackage{algorithmic}

\usepackage[caption=false,subrefformat=parens,labelformat=parens]{subfig}
\usepackage{multirow}
\graphicspath{{../figures/}}
\newcommand{\norm}[1]{\left\lVert#1\right\rVert} 
\def\BibTeX{{\rm B\kern-.05em{\sc i\kern-.025em b}\kern-.08em
    T\kern-.1667em\lower.7ex\hbox{E}\kern-.125emX}}

\begin{document}

\title{Coded Stochastic ADMM for Decentralized Consensus Optimization with Edge Computing}

\author{
	Hao~Chen,
	Yu~Ye,~\IEEEmembership{Student~Member,~IEEE,}
	Ming~Xiao,~\IEEEmembership{Senior~Member,~IEEE,}
	Mikael~Skoglund,~\IEEEmembership{Fellow,~IEEE, }
	and~H.~Vincent~Poor,~\IEEEmembership{Fellow,~IEEE }
 	\thanks{Hao Chen, Yu Ye, Ming Xiao and Mikael Skoglund are with the School of Electrical Engineering and Computer Science, Royal Institute of Technology (KTH), Stockholm, Sweden (email: haoch@kth.se, yu9@kth.se, mingx@kth.se, skoglund@kth.se).
	
	H. Vincent Poor is with Department of Electrical Engineering, Princeton University, Princeton, USA (email: poor@princeton.edu).	
 	}	
		
}
  
\maketitle
 
\begin{abstract}


Big data, including applications with high security requirements, are often collected and stored on multiple heterogeneous devices, such as mobile devices, drones and vehicles. Due to the limitations of communication costs and security requirements, it is of paramount importance to extract information in a decentralized manner instead of aggregating data to a fusion center. To train large-scale machine learning models, edge/fog computing is often leveraged as an alternative to centralized learning. We consider the problem of learning model parameters in a multi-agent system with data locally processed via distributed edge nodes.
A class of mini-batch stochastic alternating direction method of multipliers (ADMM) algorithms is explored to develop the distributed learning model. To address two main critical challenges in distributed networks, i.e., communication bottleneck and straggler nodes (nodes with slow responses), error-control-coding based stochastic incremental ADMM is investigated. Given an appropriate mini-batch size, we show that the mini-batch stochastic ADMM based method converges in a rate of $O(\frac{1}{\sqrt{k}})$, where $k$ denotes the number of iterations.
Through numerical experiments, it is revealed that the proposed algorithm is communication-efficient, rapidly responding and robust in the presence of straggler nodes compared with state of the art algorithms. 

\end{abstract}


\IEEEpeerreviewmaketitle

\section{Introduction}
\IEEEPARstart{I}{nternet} of things (IoT) devices such as mobile sensors, drones and vehicles, are widely used with emerging applications. Reference \cite{iot_device} mentions that the number of active IoT devices is expected to be over 75 billion by 2025.
The massive data generated from these devices are commonly collected and stored in a distributed manner. It is often impractical or inefficient to send all data to a centralized location due to the limitations of communication costs or latency \cite{big_data2}. Thus, data processing close to the sources or devices plays a pivotal role in avoiding high latency and communication costs. Contrary to cloud computing with centralized data processing, edge/fog computing with distributed data processing is one alternative solution to data analysis especially for large-scale machine learning models.

Commonly, machine learning with distributed computing can be formulated in the following form, among which $N$ agents cooperatively solve one optimization problem:
\begin{equation}\label{eq:main_problem1}
\min_{x}~ \sum_{i=1}^{N}f_i(x;\mathcal{D}_i),
\end{equation}
where $f_i:\mathbbm{R}^{p\times d}\to\mathbbm{R}$ is the local loss function of agent $i$, and $\mathcal{D}_i$ is the private dataset at agent $i$. The variable $x$ is shared among all agents.
Distributed machine learning has recently received growing attention from both academia and industry.
In \cite{wadmm, pwadmm, wpg, DGD, EXTRA, COCA, DADMM}, a few distributed algorithms have been developed to address optimization problem (\ref{eq:main_problem1}). Currently, primal and primal-dual methods are two main widely used solutions, which include e.g., gradient descent (GD) based methods and alternating direction method of multipliers (ADMM) based methods, respectively. In general, compared to GD, ADMM is better suited for decentralized optimization and has been demonstrated to have fast convergence in many applications, such as smart grids \cite{smart_grid}, wireless sensor networks (WSNs) \cite{wsn}, and cognitive radio networks \cite{DADMM}.

The performance of distributed consensus optimization as in (\ref{eq:main_problem1}) is commonly measured by computation time and communication costs.
In state-of-the-art approaches, agents exchange information with all, or a subset of, their one-hop neighbors. 
Existing distributed optimization schemes, such as decentralized gradient descent (DGD), and EXTRA, decentralized ADMM, Jacobi-Proximal ADMM, proposed in \cite{DGD, EXTRA, DADMM,jacobi_admm}, have good convergence rates with respect to the number of iterations (corresponding to the computation time). 
However, for large-scale machine learning problem such as distributed systems with unstable links in federated learning \cite{feder_learning}, the impact of communication costs becomes pronounced while computation is relatively cheap.
The methods in \cite{DGD, EXTRA, DADMM,jacobi_admm} are not communication efficient since multiple agents are active in parallel, and multiple communication links are used for information sharing in each iteration. 

Thus, 
alternative techniques
such as the distributed ADMM (D-ADMM) in \cite{d-admm}, the  communication-censored ADMM (COCA) in \cite{COCA}, and Group ADMM (GADMM) in \cite{gadmm}, have been proposed to limit the overall communication load in each iteration. 
Specifically, for reducing communication costs, eliminating less informative message sharing is preferred. In \cite{COCA},  the proposed COCA was able to adaptively determine whether or not a message is informative during the optimization process. Following COCA, communication-censored linearized ADMM (COLA) was introduced in \cite{cola}  to take into account hardware or time constraints in applications such as an IoT network equipped with cheap computation units or in a rapidly changing environment. 
Furthermore, an incremental learning method has also been recognized as a promising approach to reduce communication costs, which activates one agent and one link at any given time in a cyclic or a random order whilst keeping all other agents and links idle. W-ADMM \cite{wadmm}, PW-ADMM \cite{pwadmm}, and WPG \cite{wpg} are typical examples of the incremental method.
Moreover, due to the limited communication bandwidth, transmitting compressed messages via quantization \cite{qsgd, quantized_admm} or sparsification \cite{sparsified_sgd, qsparse} is also an effective method to alleviate the communication burden. Following this rationale, quantized stochastic GD (SGD) and quantized ADMM were proposed in \cite{qsgd}, and \cite{quantized_admm}, respectively. In \cite{qsparse}, the Qsparse-local-SGD algorithm was proposed, which combines aggressive sparcification with quantization and local computation along with error compensation. However, in these methods, accuracy is sacrificed to achieve lower communication costs \cite{compressed_commu}.

Apart from the communication bottleneck, the challenge of straggler nodes is also significant due to the possible presence of slow or unresponsive agents in the distributed machine learning.
To address this problem, error control coding has been applied to distributed edge computing and machine learning algorithms via computational redundancy. Coded distributed machine learning, e.g., those based on matrix multiplication \cite{speed-up}, and GD \cite{gradient_coding, jingyue, rscode}, have gained substantial research attention in various aspects. For instance, in \cite{gradient_coding}, gradient coding (GC) based on maximum distance separable (MDS) codes was first proposed to mitigate the effect of stragglers in distributed GD. In \cite{rscode}, the authors proposed a novel framework based on Reed-Solomon (RS) codes accompanied with an efficient decoder, which was used to recover the full gradient update from a fixed number of responding machines. Fountain code based schemes were developed for large-scale networks, especially when the quality of communication links was relatively low in \cite{jingyue}. 

Most existing GC schemes aim at recovering the full gradient. However, in practical large-scale distributed machine learning systems, when the amount of data is tremendously large, 
an approximate gradient via cheap unreliable nodes is more appealing as it exhibits a low computational complexity by recovering an inexact gradient in each iteration \cite{ cyclic_mds, sgc_straggler, wang2019erasurehead, ldgm_code}. 
Approximate gradient codes (AGCs) were first analyzed in \cite{cyclic_mds}.
stochastic gradient coding (SGC) was proposed for situations when the stragglers are random in \cite{sgc_straggler}. In \cite{ldgm_code}, a low density generator matrix (LDGM) code based distributed SGD scheme was proposed to recover the gradient information during the existence of slow-running machines. To the best of our knowledge, however, there is no result applying error-control coding for ADMM.

In addition, there are many research activities on mini-batch stochastic optimization in distributed settings, e.g., \cite{ouyang_admm, lian2018asynchronous, amiri2019computation, ferdinand2020anytime}. Notably,
in \cite{lian2018asynchronous}, the proposed asynchronous decentralized parallel stochastic gradient descent (AD-PSGD) enabled wait-free computation and communication. To relieve the impact of stragglers, an online distributed optimization method called Anytime Minibatch was proposed in \cite{ferdinand2020anytime}, which prevented stragglers from holding up the system without wasting the work that stragglers already completed. 
However, the relation between mini-batch size and stragglers has not been unveiled.

Motivated by these observations, we investigate decentralized learning by utilizing
ADMM as a parallel optimization tool. 
We extend our preliminary work in \cite{ye_isit} and investigate the possibility of coding for stochastic incremental ADMM (sI-ADMM) for combating straggler nodes and reducing the communication cost. The main contributions of our work can be summarized as follows:

\begin{itemize}

	\item We propose an inexact proximal stochastic incremental ADMM (sI-ADMM) to solve the decentralized consensus optimization problem, the updating order of which follows a predetermined circulant pattern. Moreover, to reduce the response time on agents, computing resources at the edge are applied to calculate partitioned gradients.  
	
	\item To provide tolerance to link failures and straggler nodes for the edge computing with ADMM, we present the coded stochastic incremental algorithm (csI-ADMM) by using coding strategies to explore the redundancy over the partitioned gradients computed by edge nodes. 

	\item The convergence and communication properties of sI-ADMM algorithms are provided through theoretical analysis and experiments. We show that our proposed csI-ADMM has a $O(\frac{1}{\sqrt{k}})$ convergence rate and $O(\frac{1}{\upsilon ^2}) $ communication. Besides, the trade-off between convergence rate and the number of straggler nodes as well as relation between mini-batch and stragglers are theoretically analyzed. Numerical results from experiments reveal that the proposed method can be communication efficient, rapidly responding and robust against the straggler nodes. 

\end{itemize}
The rest of the paper is organized as follows. 
Section \ref{system_model} presents the problem statement. 
We provide the description of the stochastic incremental ADMM algorithms in Section \ref{proposed_algorithm} and the performance analyses are presented in Section \ref{section:analysis}.
To validate the efficiency of proposed methods, we provide numerical experiments in Section \ref{sec:results}. 
Finally, we conclude the paper in Section \ref{conclusion}.

\subsection*{Notation}
Throughout the paper, we adopt the following notation: $\mathbb{E}\left[\cdot\right]$ denotes the expectation with respect to a set of variables $\bm{\xi}_i^k=\{\xi_{i,l}^k\}_{M}$. $|\cdot|$ is the absolute value. $\norm{\cdot}$ denotes the Euclidean norm $\norm{\cdot}_2$. $| \bm{\xi}_{i,j}|$ represents the cardinality of set $\bm{\xi}_{i,j}$. $\lfloor{\cdot} \rfloor$ is the floor function. $\nabla f(\cdot)$ denotes the gradient of a function $f$. $\langle \cdot, \cdot \rangle $ denotes the inner product in a finite dimensional Euclidean space. $x^*\in \mathcal{X}$ denotes the optimal solution to (\ref{eq:main_problem1}), where $\mathcal{X}$ is the domain. Besides, we define $ D_{\mathcal{X}} \overset{\Delta}{=} \mathop{ \text{sup}}\nolimits_{x_a, x_b \in \mathcal{X}} \norm{x_a - x_b}$.
    
 
\section{System Model and Problem Formulation}\label{system_model} 

\begin{figure} [t] 
	\vskip 0.2in
	\begin{center}
		\centerline{\includegraphics[width=86mm]{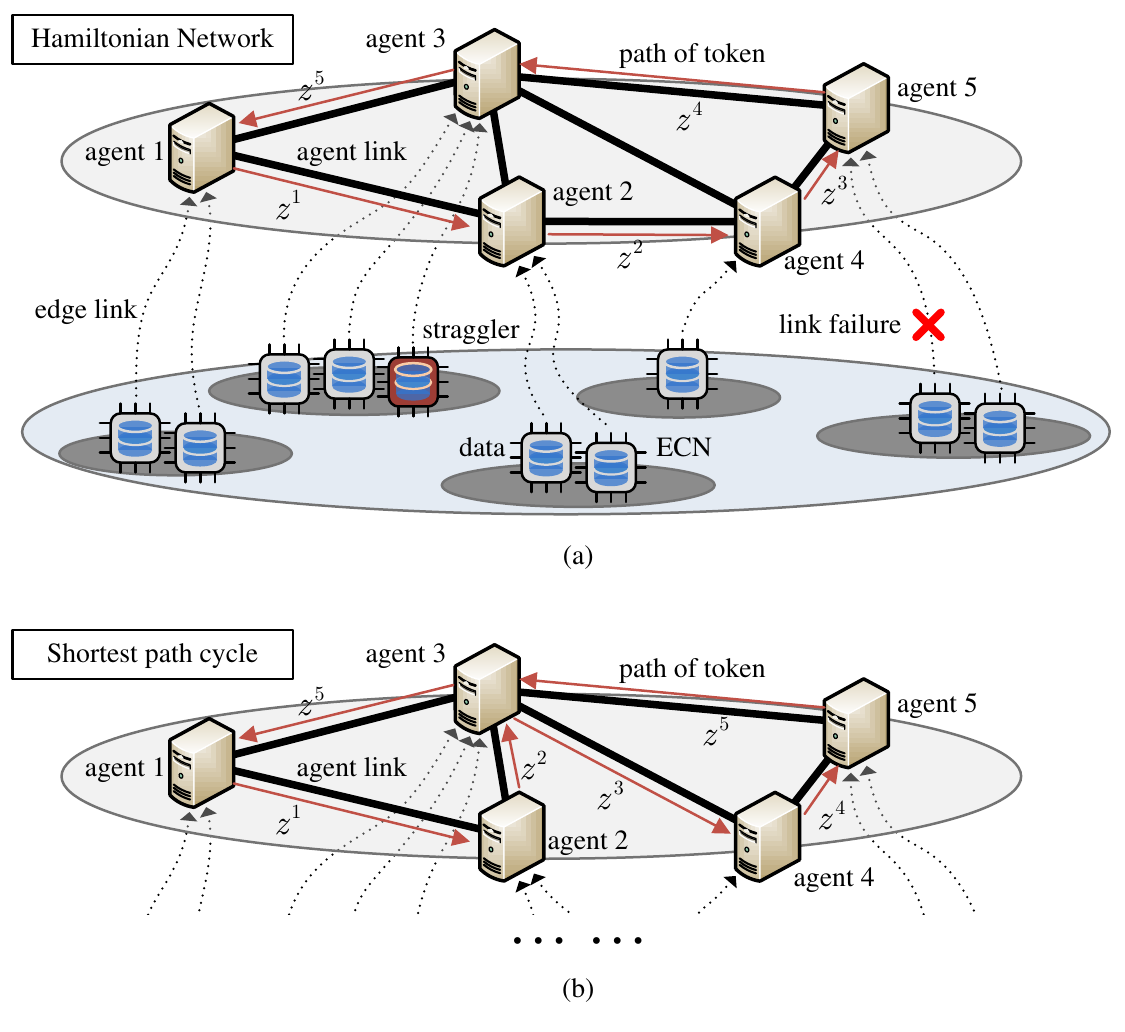}}
		\caption{Traversing patterns over different network typologies: (a) Hamiltonian network; (b) The shortest path cycle based network.}
	    \label{fig_network}
	\end{center}
	\vskip -0.2in
\end{figure}

As depicted in Fig. \ref{fig_network}, we consider a distributed computing network consisting of dispersed network elements (usually called agents in multi-agent collaborative systems) that are connected with several edge computing nodes (ECNs). Agents can communicate with each other. ECNs are capable of processing data collected from sensors, and transferring desired messages (e.g., gradient updates) back to the connected agent. 
Denote the decentralized network as $\mathcal{G}=(\mathcal{N},\mathcal{E})$, where $\mathcal{N}=\{1,...,N\}$ is the set of agents and $\mathcal{E}$ is the set of links connecting agents. 
Based on the agent coverage and computing resources, the ECNs connected to agent $i(\in \mathcal{N} )$ are denoted as $\mathcal{K}_i = \{1,...,K_i\}$.
This architecture is common in wireless sensor networks (WSNs), such as smart home systems.

We consider training a machine learning model of interest over this network, where agents can collaboratively learn a shared model parameter while keeping all the data locally.
For the agents, we make the following assumptions:
1) local views, no agent has a global view of the whole system, but behaves solely on the basis of local information (i.e., local data and model parameter); 2) decentralization, no agent controls the other agents as the master server;
and 3) one-hop communication, each agent only exchanges global model parameter information with directly connected neighboring agents. 
Multi-agent systems can obtain complex behaviors (i.e., global model) based on the interactions among agents, each of which has a simple behavior (local model).

The formulation of decentralized optimization problem can be described as follows.
The multi-agent system seeks to find out the optimal solution $x^*$ by solving (\ref{eq:main_problem1}).
$\mathcal{D}_i$ is the private dataset, which is collected from sensors such as drones and will be allocated into $K_i$ dispersed ECNs.
By defining $ \bm{x}=[x_1,...,x_N]\in\mathbbm{R}^{pN\times d}$ and introducing a global variable $z\in\mathbbm{R}^{p\times d}$, problem (\ref{eq:main_problem1}) can be reformulated as
\begin{equation}\label{eq2}
\begin{aligned}
(\text{P-1}): \min_{\bm{x}, z}~\sum_{i=1}^{N}f_i(x_i;\mathcal{D}_i),  ~~~
s.t.~  \mathbbm{1}\otimes z-\bm{x}=\bm{0},
\end{aligned}
\end{equation}
where $\mathbbm{1}=[1,...,1]^T\in\mathbbm{R}^{N}$, and $\otimes$ is the Kronecker product. In the following, $f_i(x_i,\mathcal{D}_i)$ is denoted as $f_i(x_i)$ for notational simplicity.
Our objective is to devise a both communication-efficient and straggler-tolerant decentralized algorithm such that the agents can collaboratively find an optimal solution through local computations and limited information exchange among neighbors.
In our scheme, local gradients are calculated in dispersed ECNs, while variables including primal and dual variables and global variables $z$ are updated in the corresponding agent.

\section{Proposed Stochastic ADMM algorithms} \label{proposed_algorithm}
For illustration,
we will first review the standard incremental ADMM iterations for decentralized consensus optimization. Then we present the stochastic incremental ADMM with coding in our networks.
The augmented Lagrangian function of problem (P-1) is 
\begin{equation} \label{eq:lagrangian}
\mathcal{L}_{\rho}(\bm{x},\bm{y}, z)=\sum _{i=1}^{N}f_i(x_i) + \left\langle \bm{y} ,\mathbbm{1}\otimes z-\bm{x}   \right\rangle + \frac{\rho}{2}\norm{\mathbbm{1}\otimes z-\bm{x}}^2,
\end{equation}
where $\bm{y}=[y_1,...,y_N]\in\mathbbm{R}^{pN\times d}$ is the dual variable, and $\rho>0$ is a penalty parameter. Following our preliminary work in \cite{ye_isit}, incremental ADMM (I-ADMM), with guaranteeing $\sum_{i=1}^{N}(x_i^1 - \frac{y_i^1}{\rho}) = \bm{0}$ (e.g., initialize $x_i^1 =y_i^1=\bm{0}$), the updates of $\bm{x}$, $\bm{y}$ and $z$ at the ($k+1$)$-$th iteration follow:
  \begin{subequations}
 	\begin{align}
	&x_i^{k+1}:=\left\{\begin{aligned}
	&\arg \min_{x_i}~f_i(x_i) + \frac{\rho}{2}\norm{z^k-x_i+ \frac{y_i^k}{\rho}}^2, ~i=i_k ;\\
	&x_i^{k },~\text{otherwise};
	\end{aligned}  \right. \label{old_x}\\
	&y_i^{k+1}:=\left\{\begin{aligned}
	& y_i^{k} + \rho \left( z ^{k }-x_{i}^{k +1}  \right),~i=i_k ;\\
	& y_i^{k },~\text{otherwise} ;\label{old_y}
	\end{aligned}  \right. \\
	&z ^{k+1}:=  z^{k } + \frac{1}{N}\left[  \left(x_{i_k}^{k+1}- x_{i_k}^{k } \right)
		-\frac{1}{ \rho} \left (y_{i_k}^{k+1}  - y_{i_k}^{k } \right) \right] . 
		\label{old_z} 
 	\end{align}	
 \end{subequations}

The local loss function $f_i(x_i)$ may be non-differentiable and non-convex. For ADMM, solving augmented Lagrangian especially for $x-$update above may lead to rather high computational complexity.
Approximation fitting with \textit{first-order Taylor} approximation and \textit{mini-batch stochastic} optimization will be proposed to approximate such non-linear functions and to give fast computation for $x-$update. 
To stabilize the convergence behavior of the inexact augmented Lagrangian method, a quadratic proximal term with parameter $\tau^k$ is considered.
Moreover, we also introduce the updating step-size $\gamma^k$ for the dual update.
Both parameters $\tau^k$ and $\gamma^k$ may be varying with iteration $k$.
Then, the updates of $\bm{x}$ and $\bm{y}$ at the $(k+1)$-th iteration can be presented as follows: 
  \begin{subequations}
 	\begin{align}
	&x_i^{k+1}:=\left\{\begin{aligned}
	&\arg \min_{x_i} ~\mathcal{G}_i(x_i^k;\bm{\xi}_i^k)\left(x_i-x_i^k\right) + \left\langle y_i^k,z^k-x_i \right \rangle \\ &~~~  + \frac{\rho}{2}\norm{z^k-x_i}^2 + \frac{\tau^k }{2} \norm{x_i - x_i^k}^2 , ~i=i_k ;\\
	&x_i^{k },~\text{otherwise};
	\end{aligned}  \right. \label{new_x}\\
	&y_i^{k+1}:=\left\{\begin{aligned}
	& y_i^{k} + \rho \gamma^k \left( z ^{k }-x_{i}^{k +1}  \right),~i=i_k ;\\
	& y_i^{k },~\text{otherwise} ;\label{new_y}
	\end{aligned}  \right.
 	\end{align}	
 \end{subequations}
 where $\mathcal{G}_i(x_i^k; \bm{\xi}_i^k)$ is the mini-batch stochastic gradient, which can be obtained through
 $\mathcal{G}_i(x_i^k; \bm{\xi}_i^k) = \frac{1}{M} \sum _{l=1}^{M} \nabla F_i(x_i^k; \xi_{i,l}^k)$. 
 To be more specific, $M$ is the mini-batch size of sampling data, $\bm{\xi}_i^k=\{\xi_{i,l}^k\}_{M}$ denotes a set of i.i.d. randomly selected samples in one batch and $\nabla F_i(x_i^k; \xi_{i,l}^k)$ corresponds to the stochastic gradient of a single example $\xi_{i,l}^k$.
 
\subsection{Edge Computing for Mini-Batch Stochastic I-ADMM}
We define response time as the execution time for updating all variables in each iteration.
In above updates, we assume all steps including $x $-update, $y $-update and $z$-update in agents rather than ECNs. In practice, the update is often computed in a tandem order, which leads to long response time. With the fast development of edge/fog computing, it is feasible to further reduce the response time since computing the local gradients can be dispersed to multiple edge nodes, as shown in Fig. \ref{fig_network}. 
Each ECN computes a gradient using local data and shares the result with its corresponding agent and no information is directly exchanged among ECNs.
For simplicity and analysis convenience, we focus only on the scenarios where agents are activated in a predetermined circulant pattern, e.g., according to a Hamiltonian cycle, and ECNs are activated whenever the connected agent is active, as shown in Fig.  \ref{fig_network} (a). 
A Hamiltonian cycle based activation pattern is a cyclic pattern through a graph that visits each agent exactly once (i.e., $1\rightarrow{} 2\rightarrow{}4\rightarrow{}5\rightarrow{}3$ in Fig.  \ref{fig_network} (a)).
The scenario of non-Hamiltonian cycle based traversing pattern shown in Fig. \ref{fig_network} (b), i.e., shortest path cycle based walking pattern, will be discussed in Section \ref{sec:results}.
Correspondingly, the proposed mini-batch stochastic incremental ADMM (sI-ADMM) is presented in Algorithm \ref{algorithm:1}.
At agent $i_k$, global variable $z^{k+1}$ gets updated and is passed as a token to the next agent $i_{k+1}$ via a pre-determined traversing pattern, as shown in Fig. \ref{fig_network}. 
Specifically, in the $k$-th iteration with cycle index $m=\lfloor{k/N} \rfloor$, agent $i_k$ is activated. Token $z^k$ is first received and then the active agent broadcasts the local variable $x_i^k$ to its attached ECNs $\mathcal{K}_i$. 
According to batch data with index $I_{i,j}^k$, new gradient $g_{i,j}$ is calculated in each ECN, followed by gradient update, $x $-update, $y $-update and $z$-update in agent $i_k$, via steps 20-23 in Algorithm \ref{algorithm:1}. At last, the global variable $z^{k+1}$ is passed as a token to its neighbor $i_{k+1}$.
\begin{algorithm}[t]
\caption{ Mini-batch Stochastic I-ADMM (sI-ADMM) } \label{algorithm:1}
\begin{algorithmic}[1]
	\STATE \textbf{initialize}: $\{z^1 =  x_i^1 = y_i^1=\bm{0},|i\in\mathcal{N}\}$, batch size $M$; 
	\STATE \ul{\textbf{Local Data Allocation:}}  
	\FOR{agent $i \in \mathcal{N} $}
	    \STATE \textbf{divide} $\mathcal{D}_i$ labeled data into $K_i$ equally disjoint partitions and denote each partition as $\bm{\xi}_{i,j}, j \in \mathcal{K}_i$;
	    \FOR{ECN $j  \in \mathcal{K}_i$}
	        	\STATE \textbf{allocate} $\bm{\xi}_{i,j}$ to ECN $j$;
	        	\STATE \textbf{partition} $\bm{\xi}_{i,j}$ examples into multiple batches with each size $M/K_i$;
	    \ENDFOR
	\ENDFOR
	\FOR{$k=1,2,...$}
	\STATE \ul{\textbf{Steps of Active Agent $i=i_k = (k-1)\mod N +1$:}}  
	\STATE \textbf{receive} token $z^{k }$;   
	\STATE \textbf{broadcast} local variable $x_{i}^k$ to ECNs $\mathcal{K}_i$;
	\STATE \ul{\textbf{ECN $j\in \mathcal{K}_i$ computes gradient in parallel}:}  
	\STATE \hspace{0.2cm} \textbf{receive} local primal variable $x_i^k$;
	\STATE \hspace{0.2cm} \textbf{select} batch $I_{i,j}^k= m \mod \lfloor{|\bm{\xi}_{i,j}| \cdot K_i/ M} \rfloor $;
	\STATE \hspace{0.2cm} \textbf{update} $g_{i,j}$ based on selected batch data;
	\STATE \hspace{0.2cm} \textbf{transmit} $g_{i,j}$ to the connected agent;
	\STATE \textbf{until} the $K_i$-th responded message is received;
	\STATE \textbf{update} gradient via gradient summation:
	\begin{equation}
	    \mathcal{G}_i(x_i^k; \bm{\xi}_i^k)=\frac{1}{K_i}\sum _{j=1}^{K_i} g_{i,j};
	\end{equation}
	\STATE \textbf{update} $\bm{x}^{k+1}$ according to (\ref{new_x});
	\STATE \textbf{update} $\bm{y}^{k+1}$ according to (\ref{new_y});
	\STATE \textbf{update} $z^{k+1}$ according to (\ref{old_z});
	\STATE \textbf{send} token $z^{k+1} $ to agent $i_{k+1}$ via link ($i_k, i_{k+1}$);  
	\STATE \textbf{until} the stopping criterion is satisfied.
	\ENDFOR
\end{algorithmic}  
\end{algorithm} 
\subsection{Coding Schemes for sI-ADMM}
\begin{algorithm}[t]
\caption{Coded sI-ADMM (csI-ADMM) } \label{algorithm:2}
\begin{algorithmic}[1]
	\STATE \textbf{initialize}: $\{z^1 = x_i^1 = y_i^1=\bm{0}|i\in\mathcal{N}\}$, batch size $\overline{M}$; 
    \STATE \ul{\textbf{Local Data Allocation:}}  
    \FOR{$\text{agent } i \in \mathcal{N} $}
        \STATE \textbf{divide} $\mathcal{D}_i$ labeled data based on repetition schemes in \cite{gradient_coding} and denote each partition as $\bm{\xi}_{i,j}, j \in \mathcal{K}_i$;
        \FOR{$\text{ECN } j \in \mathcal{K}_i$}
	        	\STATE \textbf{allocate} $\bm{\xi}_{i,j}$ to ECN $j$;
	        	\STATE \textbf{partition} $\bm{\xi}_{i,j}$ examples into multiple batches with each size $(S_i+1)\overline{M}/K_i$;
        \ENDFOR
    \ENDFOR 
	\FOR{$k=1,2,...$}
	\STATE  \ul{\textbf{Steps of Active Agent $i=i_k = (k-1)\mod N +1$:}}  
	\STATE \textbf{run} steps 12-13 of Algorithm \ref{algorithm:1}
	\STATE \ul{\textbf{ECN $j\in \mathcal{K}_i$ computes gradient in parallel}: } 
    \STATE \hspace{0.2cm} \textbf{run} step 15 of Algorithm \ref{algorithm:1}
	\STATE \hspace{0.2cm} \textbf{select} batch 
	\begin{equation}
	    I_{i,j}^k= m \mod\lfloor{|\bm{\xi}_{i,j}| \cdot K_i/ (S_i+1)\overline{M} } \rfloor  ;
	\end{equation}
	\STATE \hspace{0.2cm} \textbf{update} $g_{i,j}$ via encoding function $p_{enc}^j(\cdot)$;
	\STATE \hspace{0.2cm} \textbf{transmit} $g_{i,j}$ to the connected agent;
	\STATE \textbf{until} the $R_i$-th fast responded message is received;
	\STATE \textbf{update} gradient via decoding function $q_{dec}^i(\cdot)$;
	\STATE \textbf{run} steps 21-25 of Algorithm \ref{algorithm:1};
	\ENDFOR 
\end{algorithmic} 
\end{algorithm}
With less reliable and limited computing capability of ECNs, straggling nodes may be a significant performance bottleneck in the case of learning networks. To address this problem, error control codes have been proposed to mitigate the impact of the straggling nodes without knowing their locations by leveraging data redundancy.
One type of optimal linear codes, $(n,k)$ MDS codes $(k< n)$ are proposed to combat stragglers, which have $n$ coded blocks such that all $k$ message blocks can be reconstructed from any $k$ coded blocks.
Following the work in \cite{gradient_coding}, two MDS-based coding methods  over real field $\mathbbm{R}$,
i.e., \textit{Fractional} repetition scheme and \textit{Cyclic} repetition scheme are adopted and integrated with sI-ADMM for reducing the responding time in the presence of straggling nodes. The details for the two schemes can be found in \cite{gradient_coding}.
We formally present the proposed coded sI-ADMM (csI-ADMM) in Algorithm \ref{algorithm:2}.
Different from Algorithm \ref{algorithm:1}, in Algorithm \ref{algorithm:2}, encoding and decoding processes are introduced to calculate $\mathcal{G}_i(x_i^k; \bm{\xi}_i^k)$ via computation redundancy, thereby reducing the impact of straggling ECNs.
Denoting $R_i$ as the minimum required ECNs number, each agent $i$ updates local variables with the updated gradient from any $R_i$ out of $K_i$ ECNs per iteration to combat slow links and straggler nodes from stalling overall computation. 
Thus, agent $i$ can tolerate $\mathbf{any}$ $(S_i=K_i-R_i)$ stragglers.

\begin{figure} [b] 
	\vskip 0 in
	\begin{center}
		\centerline{\includegraphics[width=89mm]{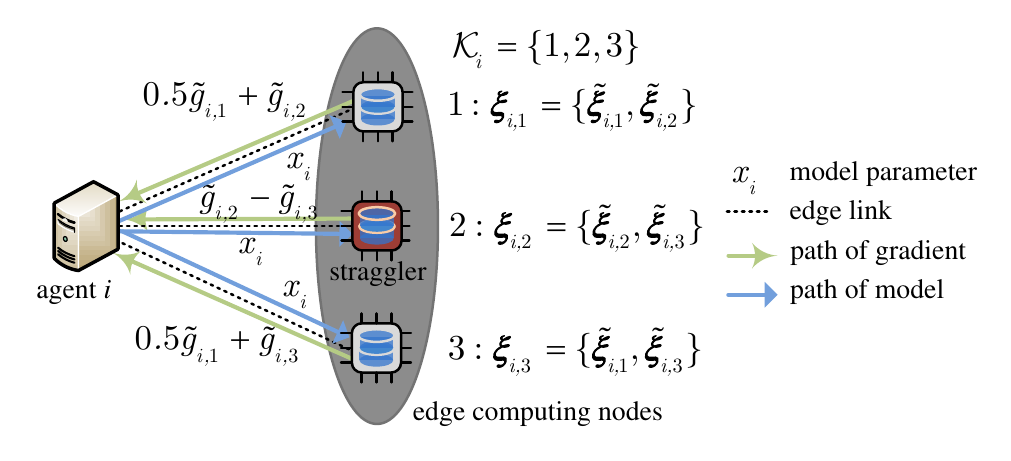}}
		\caption{Coded edge computing for mitigating the straggler nodes.}
		\label{fig:coded_edge_compute}
	\end{center}
	\vskip 0 in
\end{figure}

Fig. \ref{fig:coded_edge_compute} illustrates an example on how coded edge computing can reduce response time.
Here we assume at most $S_i=1$ ECN may be slow during each iteration (e.g., ECN $2$).
The extension to multiple stragglers is straightforward.
In Fig. \ref{fig:coded_edge_compute}, three ECNs have overlapped labeled data allocated privately (i.e., $\tilde{\bm \xi}_{i,1}, \tilde{\bm \xi}_{i,2} \text{ and }  \tilde{\bm {\xi}}_{i,3}$) and share local primal variable $x_i$. For the gradient update, once agent $i$ is activated, current local variable $x_i$ is broadcasted to all connected ECNs. ECN $1$ calculates the gradients of the shared model with $\tilde{\bm \xi}_{i,1} \text{ and } \tilde{\bm \xi}_{i,2}$ separately, where the corresponding gradients are denoted as $\tilde g_{i,1}$ and $\tilde g_{i,2}$, respectively. Denoting $\tilde {\bm  g}_i = [\tilde g_{i,1},\tilde g_{i,2},\tilde g_{i,3}]$ and  following the $(K_i, R_i)$ MDS codes in \cite{gradient_coding}, ECN~1 then computes the encoded gradient $g_{i,1}=p_{enc}^1(\tilde {\bm  g}_i)=\frac{1}{2}\tilde g_{i,1} + \tilde g_{i,2}$, i.e., the linear combination of $\tilde g_{i,1}$ and $\tilde g_{i,2}$. Similarly, ECNs 2 and 3 compute $g_{i,2}=p_{enc}^2(\tilde {\bm  g}_i) = \tilde g_{i,2} - \tilde g_{i,3}$ and $g_{i,3}=p_{enc}^3(\tilde {\bm  g}_i) = \frac{1}{2}\tilde g_{i,1} + \tilde g_{i,3}$, respectively. Then three coded gradients denoted by $\bm g_i=[g_{i,1},g_{i,2},g_{i,3}]$ are transmitted back to the agent, where any of first two arrived messages can recover the summation $\tilde g_{i,1}+ \tilde g_{i,2}+\tilde g_{i,3}$ (i.e., $\mathcal{G}_i(x_i^k;\bm{\xi}_i^k)$) through decoding function $q_{dec}^i(\bm g_i)$ followed by $x$-update, $y$-update and $z$-update for optimizing problem (P-1). Thus, comparing with uncoded learning schemes with labeled data disjointedly allocated (i.e., sI-ADMM), the response time for updating local and global variables is only decided by the first two fast ECNs instead of the slowest ECN. This may cumulatively reduce the total running time significantly. 
 
\section{Algorithm Analyses}\label{section:analysis}
In this section, we will first provide the convergence properties for both sI-ADMM and csI-ADMM, in terms of the convergence rate. Then we also analyze communication cost, defined as the amount of communication among agents and the impact of straggling nodes. In the following analysis, communication of variables between a pair of agents is taken as 1 unit of communication.
\subsection{Convergence Analysis}
We first analyze the convergence properties for the proposed algorithms.
Without loss of generality, the updating order for the proposed sI-ADMM based algorithm follows a predetermined pattern, i.e., Hamiltonian cycle order: $1\rightarrow{} 2\rightarrow{}...\rightarrow{}N\rightarrow{}1\rightarrow{}2...$ as I-ADMM in \cite{ye_isit}. 

To establish the convergence for sI-ADMM algorithms, we first make the following assumptions, which are commonly employed in the analysis of stochastic optimization. 

\begin{assumption}[Connectivity] \label{assump1}
	The graph $\mathcal{G}$ is connected and there exists at least one Hamiltonian cycle.
\end{assumption}

\begin{assumption}[Lipschitz continuous gradient] \label{assump2}
The local loss function $f_i({x})$ is lower bounded over ${x}$, and is coercive over $x$, i.e., $f_i(x)\to \infty$ if $x\in\mathcal{X}$ and $\|x \|\to \infty$.
$f_{i}(x)$ is $L$-Lipschitz differentiable, i.e., for any $x,y \in\mathbbm{R}^{p\times d}$,
\begin{equation}
\begin{aligned}
\norm{ \nabla f_{i}(x) -  \nabla f_{i}(y) } \leqslant L \norm{  x -y  }, \forall i \in\mathcal{N},
\end{aligned}
\end{equation}
and this is equivalent to 
\begin{equation}
f_i(x) \leqslant f_i(y) + \left\langle  \nabla f_i(y), x-y  \right\rangle + \frac{L}{2} \norm{x-y}^2.
\end{equation}
\end{assumption}


\begin{assumption}[Unbiased estimation]\label{assump:unbias_g}
For the differential function $f_{i}(x)$, there exists a stochastic first-order oracle that returns a noisy estimate of the gradient of $f_{i}(x)$, and the unbiased estimate $\nabla F_{i}(x, \xi_{i,l})$ satisfies
\begin{equation}
	\begin{aligned}
	\mathbbm{E}_{\xi_{i,l} \in \mathcal{D}_i} \big[\nabla F_{i}(x; \xi_{i,l})\big] = \nabla f_i(x).
	\end{aligned}
\end{equation}


Let $M$ be the size of mini-batch $\bm{\xi}_i$, i.e., $|\bm{\xi}_i|=M$, and $\bm{\xi}_{i} = \{\xi_{i,1},...,\xi_{i,M}\}\subseteq \mathcal{D}_i$ denotes a set of i.i.d. random variables, and the mini-batch stochastic gradient is given by
\begin{equation}
\mathcal{G}_i(x, \bm{\xi}_i) = \frac{1}{M} \sum _{l=1}^{M} \nabla F_i(x, \xi_{i,l}).
\end{equation}
Clearly, we have
\begin{equation}
\begin{aligned}
\mathbbm{E}_{\bm \xi_i  \sim \mathcal{D}_i}\big[ \mathcal{G}_i(x, \bm \xi_i )\big]= \nabla f_i(x).
\end{aligned}
\end{equation}


\end{assumption}
Then, with these assumptions, we first present the following convergence properties for algorithm sI-ADMM.
\begin{theorem}[Convergence]\label{theorem_1}
 Under $Assumptions$ \ref{assump1}-\ref{assump:unbias_g}, with $\{ \gamma^k \geq 4N,~\tau^k \geq \frac{2\rho}{\gamma^k } + \frac{L}{2} - \frac{\rho}{2}|k\geq 1 \}$, iterates $(\bm{x}^k, \bm{y}^k, z^k)$ generated by sI-ADMM satisfy the following properties: 
	
\begin{enumerate} 
   \item $ \mathbbm{E}\big[ \mathcal{L}_{\rho}(\bm{x}^{k+1},\bm{y}^{k+1},z^k) - \mathcal{L}_{\rho}(\bm{x}^{k+1},\bm{y}^{k+1},z^{k+1})\big]$\\
      $=\frac{N\rho}{2}\left\|z^{k} - z^{k+1} \right\|$;\label{statement1} 
      
    \item $\mathbbm{E}\big[\mathcal{L}_{\rho}(\bm{x}^k,\bm{y}^k,z^k) - \mathcal{L}_{\rho}(\bm{x}^{k+1},\bm{y}^{k+1},z^k)\big]\\
    \geq-\frac{1}{\rho \gamma^k  } \norm{\bm{y}^{k+1} - \bm{y}^k}^2 +  \big(\frac{\rho - L + 2\tau^k}{2}\big)\norm{\bm{x}^{k+1} -\bm{x}^k }^2$;\label{statement2} 
      
	\item $\mathbbm{E}\big[\mathcal{L}_{\rho}(\bm{x}^k,\bm{y}^k,z^k)-\mathcal{L}_{\rho}(\bm{x}^{k+1},\bm{y}^{k+1},z^{k+1})\big] \\
	\geq \big(\frac{\rho - L + 2\tau^k}{2} - \frac{2\rho}{\gamma^k  }\big) \norm{ \bm{x}^{k+1} -\bm{x}^k }^2\\
	~~~+ \frac{\rho N (\gamma^k - 4N)}{2\gamma^k  } \norm{z^{k+1} - z^k}^2$;\label{statement3} 
	
	\item $ \{ \mathbbm{E}\big[\mathcal{L}_{\rho}(\bm{x}^k,\bm{y}^k,z^k)\big] \}_{k\geq1}  $ is lower bounded.\label{statement4} 
\end{enumerate}
\end{theorem}
Hence, with statements $1)-4)$, the sequence $\{\mathbbm{E}\big[\mathcal{L}_{\rho}(\bm{x}^k,\bm{y}^k,z^k)\big] \}_{k\geq1}$ is convergent.
\begin{IEEEproof}
The convergence proof of sI-ADMM is similar to that of I-ADMM in \cite{ye_isit}. By substituting equation (25) of Lemma 4 in \cite{ye_isit} with $       \mathbbm{E}\big[\mathcal{G}_{i_k}(x_{i_k}^{k};\bm \xi_{i_k}^k ) - y_{i_k}^k\big] =\nabla f_{i_k}(x_{i_k}^{k}) - y_{i_k}^k = \rho(z^k-x_{i_k}^{k+1}) - \tau^k (x_{i_k}^{k+1} - x_{i_k}^k)   = \frac{1}{\gamma^k} (y_{i_k}^{k+1} - y_{i_k}^k) - \tau^k (x_{i_k}^{k+1} - x_{i_k}^k )$, Theorem \ref{theorem_1} can be obtained.

\end{IEEEproof}
We note that Theorem \ref{theorem_1} provides a sufficient condition to guarantee the convergence of the proposed sI-ADMM. 
csI-ADMM has the same convergence properties as those of sI-ADMM. 
To obtain the convergence rate for the proposed algorithms, we introduce two more assumptions as follows. 
\begin{assumption}[Bounded gradient and variance]\label{assump:bounded}
The gradient of local loss function $f_i(x)$ is bounded. That is, there exists a constant $\phi$ such that for all $x$,
\begin{equation}
\begin{aligned}
\max_{1\leqslant i \leqslant N} \text{sup}_{x \in \mathcal{X} } \norm{ \nabla f_i(x)}^2 \leqslant \phi.
\end{aligned}
\end{equation}
Moreover,
\begin{equation}
\begin{aligned}
\mathbbm{E}_{\xi_{i,l} \in \mathcal{D}_i} \left[\norm{ \nabla F_{i}(x; \xi_{i,l}) - \nabla f_i(x)   }^2\right] \leqslant \delta^2,
\end{aligned}
\end{equation}
and
\begin{equation}
\begin{aligned}
\mathbbm{E}_{\bm{\xi_i} \sim \mathcal{D}_i}\left [\norm{ \mathcal{G}_i(x, \bm \xi_i ) - \nabla f_i(x)   }^2\right] \leqslant \frac{\delta^2}{M}.
\end{aligned}
\end{equation}
\end{assumption}

\begin{assumption}[Strong convexity]\label{assump:s_c}
The local loss function $f_i(x)$ is $\mu-strongly$ convex, satisfying that
\begin{equation}
    f_i(x) \geq f_i(y) + \left\langle  \nabla f_i(y), x-y  \right\rangle + \frac{\mu}{2} \norm{x-y}^2.
\end{equation}

\end{assumption}

 
Then we conclude the convergence rate of sI-ADMM algorithm as follows.
\begin{theorem}[Convergence rate]\label{theorem1}
For $ k=mN + i$ where cycle index $m=0,...,T-1$ and $ i \in\{1,...,N\}$, taking $\tau^k = c_{\tau} \sqrt{k}, \gamma^k = \frac{c_{\gamma}}{\sqrt{k}}$ with constants $c_{\tau}, c_{\gamma} >0$ in Algorithm \ref{algorithm:1}, under Assumptions \ref{assump1}-\ref{assump:s_c} with $\beta > 0$, we obtain the following convergence rate
\begin{equation} \label{eq:theorem}
    \begin{aligned}
        & \frac{1} {TN}  \sum_{m=0}^{T-1} \sum_{i=1}^{N}\mathbbm{E} \left[f_{i}(x_{i}^{k+1})- f_{i}(x_{i}^*)\right]    \\&\qquad\qquad\qquad + \beta \mathbbm{E} \left[\norm{ \frac{1}{TN} \sum_{m=0}^{T-1} \sum_{i=1}^{N} \left(z^k - x_{i}^{k+1}\right)  } \right]\\
        &\qquad\leq \frac{1}{\sqrt{TN}}\left( \frac{c_{\tau}ND_{\mathcal{X}}^2 }{2 }  + \frac{2N\beta^2}{\rho c_{\gamma} }  + 2\phi   + \frac{\delta ^2 }{M  }\right),
    \end{aligned}
\end{equation}
if the total number of cycles $T$ is sufficiently large (i.e., iteration number $k$ is sufficiently large), especially with constraints
\begin{equation}
    \begin{aligned}
        &\mu > 3\rho, ~c_{\tau} > \frac{2}{(N+1){N}},~ \frac{1}{\mu - 3\rho} < c_{\gamma} < \frac{1}{\rho}.
    \end{aligned}
\end{equation}
\end{theorem}
\begin{IEEEproof}
The proof is relegated in Appendix \ref{secondAppendix}.
\end{IEEEproof}

\begin{remark}
Theorem \ref{theorem1} suggests that the sub-linear convergence rate for sI-ADMM algorithm is $O(\frac{1}{\sqrt{k}})$. With the network size $N$ scaling up, it indicates that the convergence rate of sI-ADMM may decrease to $O(\frac{N}{\sqrt{k}})$. Batch size $M$ plays a small role in determining the overall convergence speed although a larger batch size promotes faster convergence. Besides, this rate is also determined by the variance of stochastic gradients.
\end{remark}
\subsection{Communication Analysis}
Next, we analyze the communication cost based on the sub-linear convergence rate.
\begin{corollary}[Communication cost] \label{corollary}
Let $c_{\tau} = \frac{1}{N}$, $ c_{\gamma} = N$ and $ k=mN + i$ where $ m =  0,...,T-1 , i \in \{1,...,N\}$, under the same conditions as those in Theorem \ref{theorem1}, with mean deviation defined by 
\begin{equation}\label{eq12}
    \frac{1 }{TN}  \sum_{m=0}^{T-1} \sum_{i=1}^{N} \mathbbm{E}       \left[\norm{{f_{i}(x_{i}^{k+1})- f_{i}(x_{i}^*)} } \right]  \leq \upsilon,
\end{equation} 
the communication cost of the proposed sI-ADMM is $O(\frac{1}{\upsilon ^2})$. 
\end{corollary}
\begin{IEEEproof}
From (\ref{eq:theorem}) with $c_{\tau} = \frac{1}{N}, c_{\gamma} = N$, to achieve mean deviation (\ref{eq12}), it is enough to have
\begin{equation}
   \frac{1}{\sqrt{TN}}  \left(\frac{D_{\mathcal{X}}^2}{2}   + \frac{2\beta^2}{\rho }  +2\phi + \frac{\delta ^2 }{M} \right)\leq \upsilon,
\end{equation}
which is implied by
\begin{equation}
    k = TN \geq \frac{1}{\upsilon ^2} \left(\frac{D_{\mathcal{X}}^2 }{2}  + \frac{2\beta^2}{ \rho }  +2\phi + \frac{\delta ^2 }{M}  \right)^2.
\end{equation}
For each cycle $m$, there are $N$ iterations, which has $O(N)$ communication.
Since $\frac{D_{\mathcal{X}}^2 }{2}  + \frac{2\beta^2}{\rho }  +2\phi + \frac{\delta ^2 }{M}  $ can be regarded as constant with respect to network size (i.e., agent number $N$), to guarantee (\ref{eq12}), the communication cost is $O(\frac{1}{\upsilon ^2}) $.
This completes the proof.
\end{IEEEproof}

For csI-ADMM algorithm, both communication cost and convergence rate are roughly the same as sI-ADMM (but with differences outline in sub-Section \ref{sub-sec:impact} below).

\subsection{The Impact of Straggling Nodes} \label{sub-sec:impact}
From Theorem \ref{theorem1}, it is proven that for the proposed (c)sI-ADMM, a faster convergence speed can be achieved with a larger mini-batch size setup. However, in order for the csI-ADMM algorithm to tolerate more straggler nodes, the ECNs' capacity such as memory and storage, limits the maximum allowable mini-batch size per iteration. Under the same condition of computation overhead in ECNs, if pursuing robust to more stragglers, more overlapped data, i.e., less disjoint data, are participated in combating with stragglers in each iteration.
The trade-off between the number of tolerated straggler nodes and mini-batch size can be formulated as
\begin{equation}
    \overline{M} = \frac{M}{S+1},
    \label{eq:s_vs_batchsize}
\end{equation}
where $M$ is the selected mini-batch size for the case without straggler nodes, $\overline{M}$ is the maximum potential mini-batch size for the case with $S$ straggler nodes.
\begin{corollary}[Convergence rate] \label{corollary_2}
Under the conditions of Theorem \ref{theorem1}, suppose that there exists $S_i=S$ straggling ECNs connected to each agent $i$. 
The coded csI-ADMM algorithm roughly achieves a $O( \frac{1}{\sqrt{k}}\cdot\frac{S+M+1}{M})$ convergence rate.
\end{corollary}
\begin{IEEEproof}
By substituting $\overline{M}$ into (\ref{eq:theorem}), we can obtain the desired result.
\end{IEEEproof}
 
Apparently, Corollary \ref{corollary_2} implies that, to combat with more straggler nodes, the allowed batch size $\overline{M}$ is smaller than the case with robustness to less straggling ECNs.
However, smaller batch size degrades the algorithm convergence speed. 
In the following Section \ref{sec:results}, the relation between the number of allowed straggler nodes and convergence rate will be also verified through numerical experiments.
As for the response time for both uncoded and coded distributed systems, the relevant analysis can be found in \cite{speed-up}. In practice, the system cannot wait long time for the slowest ECN. Hence, a maximum delay parameter $\epsilon$ will be considered in following experiments.


\section{Numerical Experiments} \label{sec:results}
\begin{figure*}[t] 
 	\centering
 	\vskip -0.2in
	\subfloat[ ]{\hspace*{1mm}\includegraphics[width=60mm]{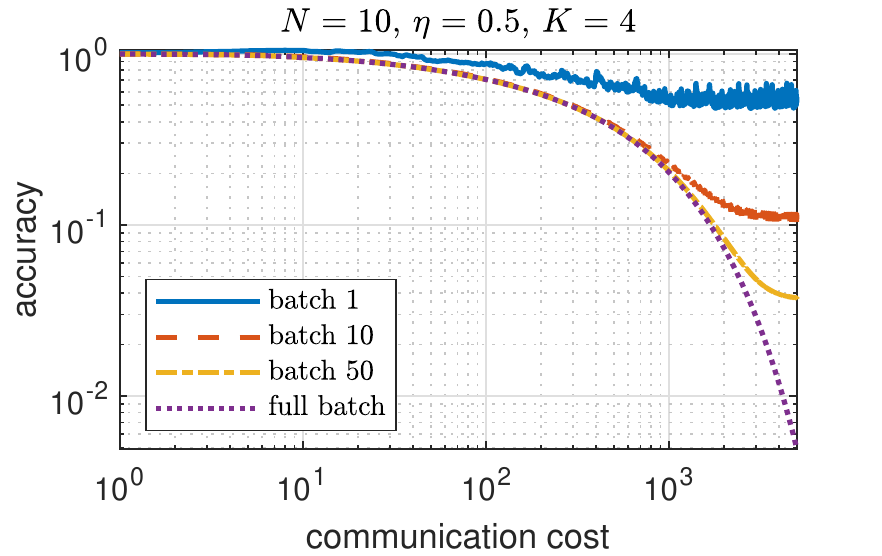}\label{fig_batch_a}}
	\subfloat[ ]{\hspace*{-2mm}\includegraphics[width=60mm]{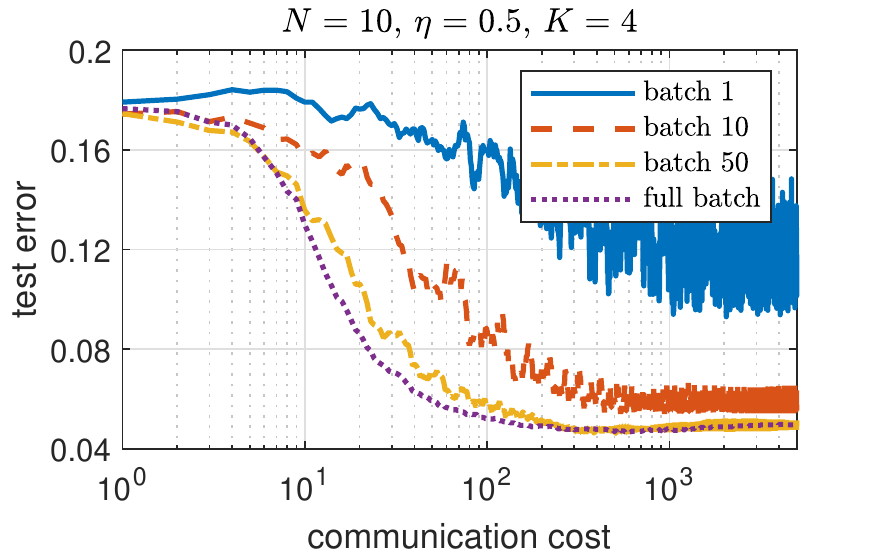}\label{fig_batch_b}}
 	\subfloat[ ]{\hspace*{-2mm}\includegraphics[width=60mm]{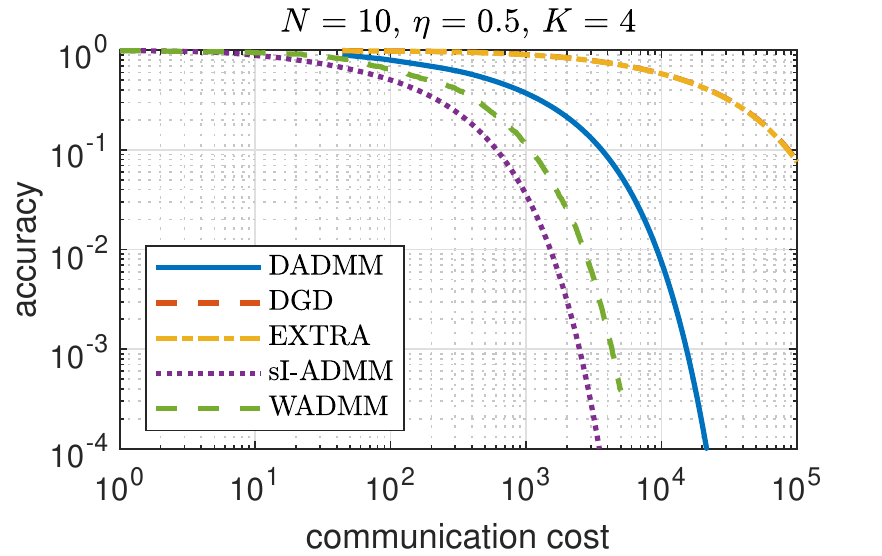}\label{fig_compare_c}} \\ 
    \subfloat[ ]{\hspace*{1mm}\includegraphics[width=60mm]{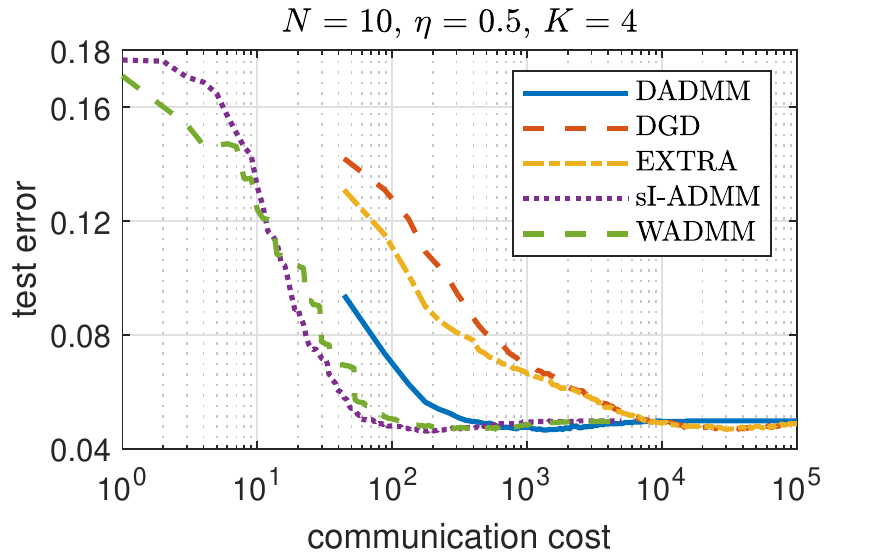}\label{fig_compare_d}}
 	\subfloat[ ]{\hspace*{-2mm}\includegraphics[width=60mm]{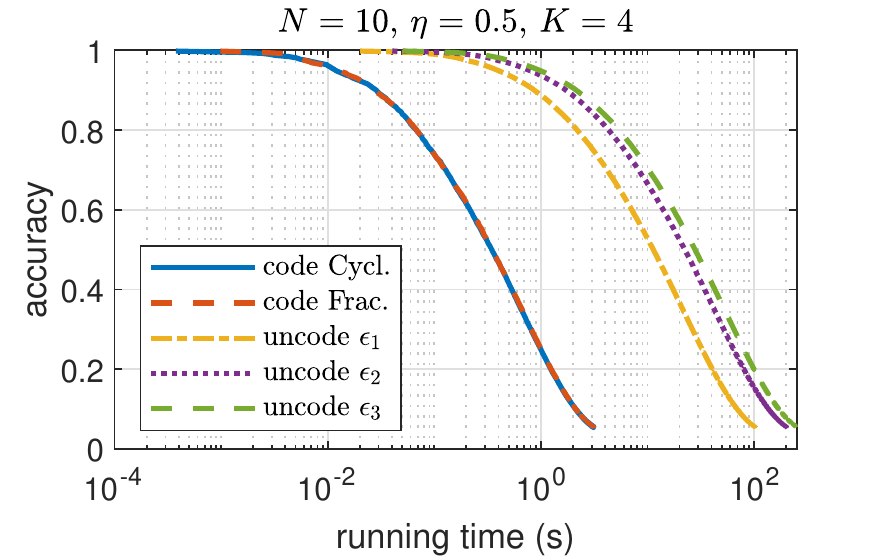}\label{fig_siadmm_e}} 
 	\subfloat[ ]{\hspace*{-2mm}\includegraphics[width=60mm]{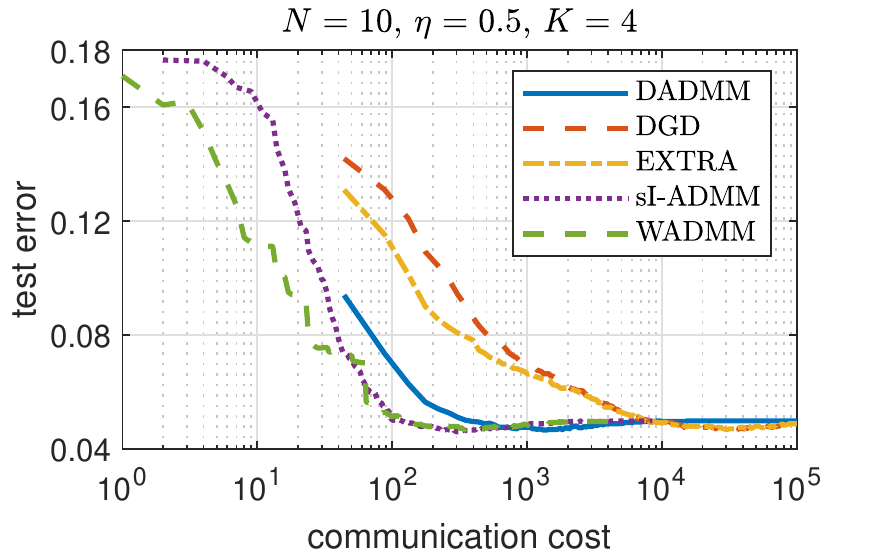}\label{fig_compare_g}} 
 	\vskip -0.1in
 	\caption{Performance of different consensus optimization methods on least squares on dataset USPS.}
 	\label{fig_performance_usps}
 \end{figure*}

In this section, both synthetic and real-world datasets are tested to evaluate the performance of the proposed stochastic ADMM algorithms for decentralized consensus optimization.
We evaluate convergence performance in terms of mini-batch size, communication cost, running time as well as number of straggler nodes.

\subsection{Simulation Setup}
The experimental network $\mathcal{G}$ consists of $N$ agents and $E = \frac{N(N-1)}{2}\eta$ links, where $\eta$ is the network connectivity ratio.
For agent $i$, $K_i = K$ ECNs with the same computing power (e.g., computing and memory) are attached.
To reduce the impact of token traversing patterns, both Hamiltonian cycle-based and non-Hamiltonian cycle-based (i.e., the shortest path cycle-based \cite{wpg}) token traversing methods are evaluated for the proposed algorithms. 
For the shortest path cycle based traversing method, Fig. \ref{fig_network} (b) illustrates the token traversing pattern.
The traversing route is determined through the shortest path routing strategy \cite{sp-cycle} and the cycle is formed through concatenating multiple shortest paths.

To investigate the communication efficiency, we compare our approaches with state-of-the-art consensus optimization methods: 1) WADMM in \cite{wadmm}, where the agent activating order follows a random walk over the network, 2) D-ADMM in \cite{d-admm}, 3) DGD in \cite{DGD} and 4) EXTRA in \cite{EXTRA}, with respect to the relative error, which is defined as
\begin{equation}
\text{accuracy} = \frac{1}{N} \sum_{i=1}^{N} \frac{\norm{x_i^k - x^*}}{\norm{x_i^1 - x^*}},
\end{equation}
where $x^* \in\mathbbm{R}^{p\times d}$ is the optimal solution of (P-1). 
For demonstrating the robustness against straggler nodes, distributed schemes, including \textit{Cyclic} and \textit{Fractional} repetition methods and uncode method, are achieved for comparison. For fair comparison, the parameters for algorithms are tuned, and kept the same in different experiments. Moreover, unicast is considered among agents and the communication cost per link is 1 unit. The consumed time for each communication among agents is assumed to follow a uniform distribution $\mathcal{U}(10^{-5}, 10^{-4})$ s. The response time of each ECN is measured by the computation time and the overall response time of each iteration is equal to the execution time for updating all variables in each iteration. 
Moreover, a maximum delay $\epsilon$ for stragglers in each iteration is considered in simulation. 
All experiments were performed using Python on an Intel CPU @2.3GHz (16GB RAM) laptop.

\begin{figure*}[t] 
 	\centering
 	\vskip -0.1in
	\subfloat[ ]{\hspace*{1mm}\includegraphics[width=60mm]{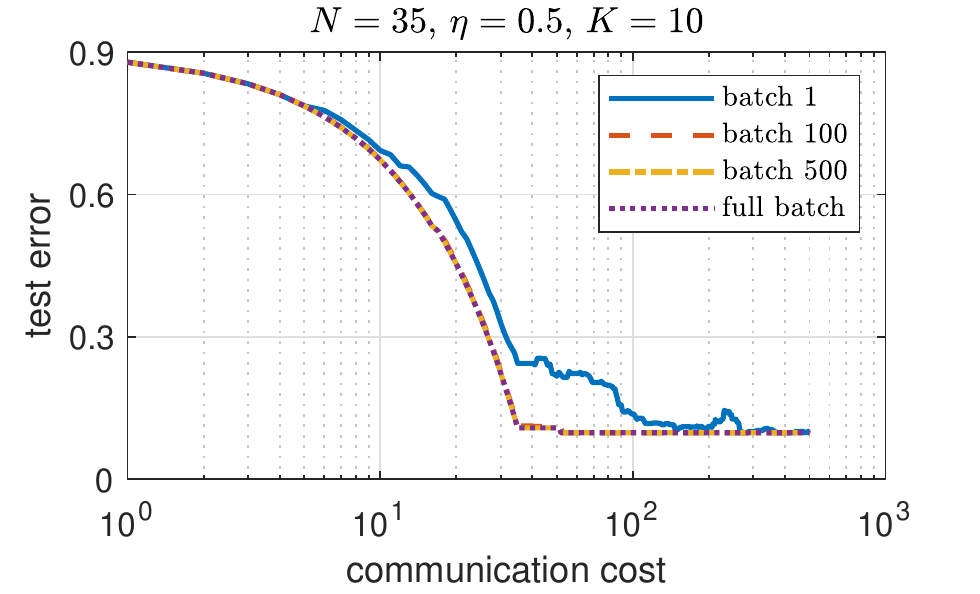}\label{fig5_batch_a}}
	\subfloat[ ]{\hspace*{-2mm}\includegraphics[width=60mm]{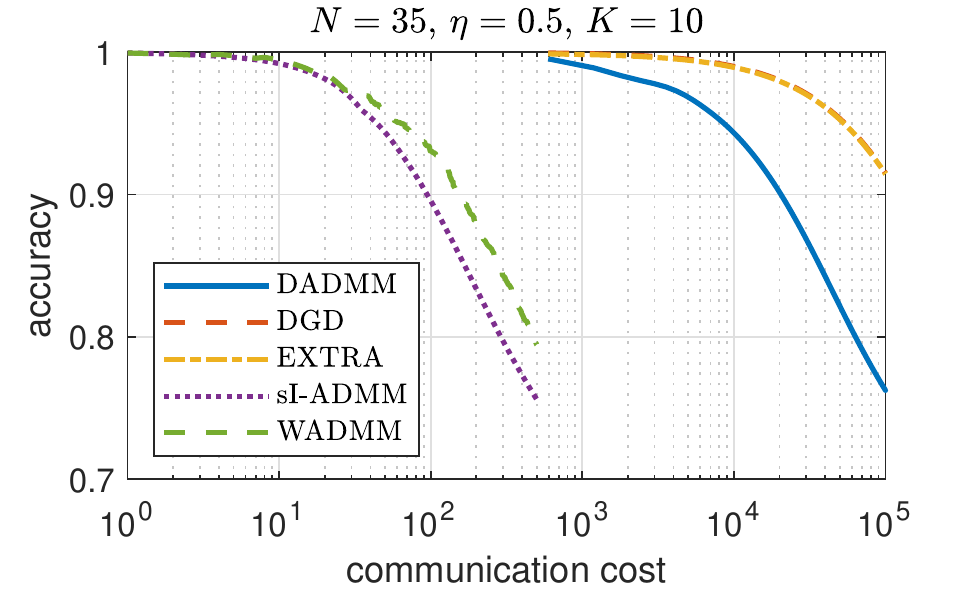}\label{fig5_compare_b}}
 	\subfloat[ ]{\hspace*{-2mm}\includegraphics[width=60mm]{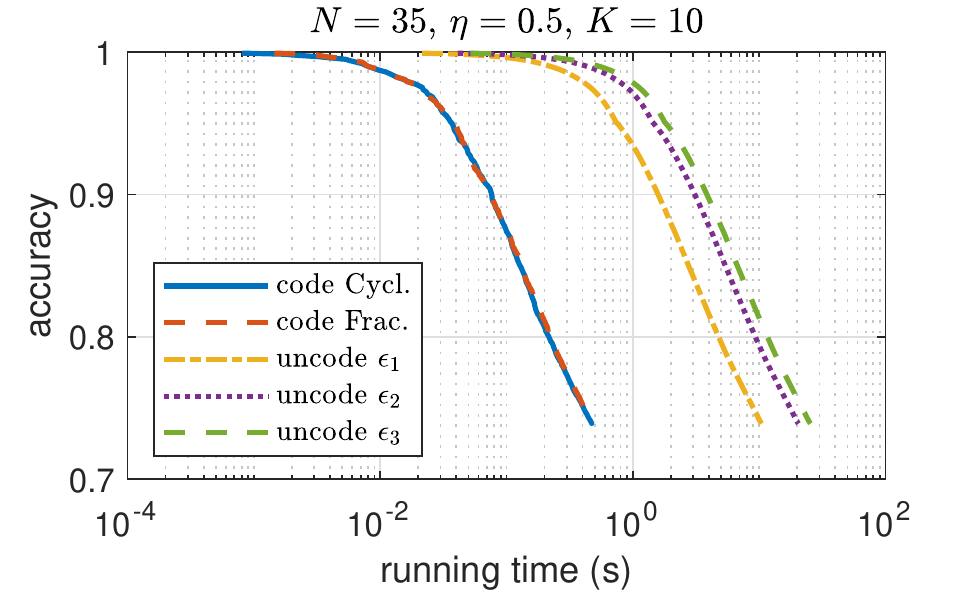}\label{fig5_siadmm_c}}

    \caption{Performance of different consensus optimization methods on least squares on dataset ijcnn1.}
    \label{fig_performance_ijcnn1}
 \end{figure*}
 
We consider the decentralized least square problem, which aims at solving (\ref{eq:main_problem1}) with the local function of each agent  
\begin{equation}
f_i(x_i,\mathcal{D}_i) = \frac{1}{2b_i}\sum_{j=1}^{b_i}\norm{x_i^T o_{i,j} - t_{i,j}}^2,
\end{equation}
where $\mathcal{D}_i = \{ o_{i,j}, t_{i,j} |j=1,...,b_i \}$ is the total disjoint dataset that agent $i$ needs to allocate among $K_i$ ECNs. 
In simulation, both synthetic and real datasets are utilized, which are summarised in Table \ref{tab:dataset}. 
For synthetic dataset, entries of $x_o \in \mathbbm{R}^{3 \times 1}$ and input $o_{i} \in \mathbbm{R}^3$ are generated with independent standard normal distribution. 
Output measurement $t_{i} \in \mathbbm{R}^1$ follows $t_{i} := x_o^T o_{i} + e_{i} $, where $e_{i} \sim \mathcal{N}(0, \sigma I_1)$ is the random noise with variance $\sigma$.
Both USPS and ijccn1 data are disjointly linked to all agents. 
And among $K_i$ ECNs, we divide all the local data $D_i$ equally and disjointly, assigning $(S_i +1)$ partitions to each ECN.
\begin{table}[!h]
    \caption{Simulation Datasets for Decentralized Consensus Optimization}
    \label{tab:dataset}
    \centering
    \fontsize{9}{8}\selectfont
    \begin{tabular}{|c|c|c|c|c|}
        \hline 
        datasets & \# training & \# test & \# Dim. (p) & \# Dim. (d) \\
        \hline 
        synthetic & 50,400 & 5,040 & 3 & 1 \\ \hline
        USPS \cite{usps} & 1,000 & 100 & 64 & 10 \\ \hline
        ijcnn1\cite{libsvm} & 35,000 & 3,500 & 22 & 2\\ 
        \hline
    \end{tabular}
\end{table}

\subsection{Simulation Results}
Fig. \ref{fig_performance_usps} and \ref{fig_performance_ijcnn1} show the convergence performance of different consensus optimization methods on the least squares using dataset USPS, and dataset ijcnn1, respectively. 
Specifically, for Fig. \ref{fig_performance_usps}, a test network with Hamiltonian cycle is first considered in sub-figures Fig. \ref{fig_performance_usps} (a)-(e), while one result with the shortest path-based cycle is shown in Fig. \ref{fig_performance_usps} (f).
In sub-figures Fig. \ref{fig_performance_usps} (a) and (b), we present the impact of $M$, the size of mini-batch, on the convergence behavior. It can be concluded that with increasing $M$, a higher accuracy of the proposed algorithms can be achieved with the same communication cost whilst the test error is lower as well.
This agrees with Theorem \ref{theorem1} that a large mini-batch size may lead to fast convergence.
The accuracy vs. communication cost and the test error vs. communication cost are shown in sub-figures Fig. \ref{fig_performance_usps} (c), (d), and (f), where test error is defined as the mean square error loss. 
 \begin{figure}[t] 
 	\centering
 	\vskip -0.1in
	\subfloat[ ]{\hspace*{1mm}\includegraphics[width=45mm]{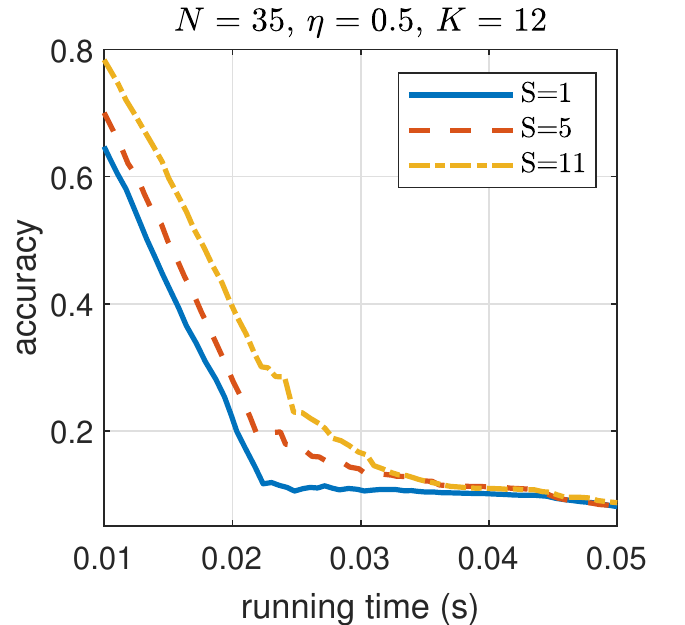}\label{fig_ham1}}
	\subfloat[ ]{\hspace*{-2mm}\includegraphics[width=45mm]{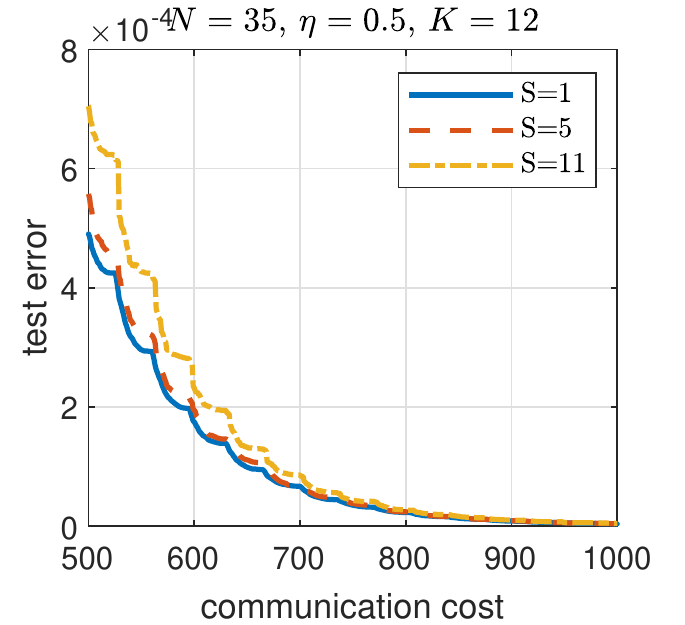}\label{fig_b1}}
    \caption{Impact of number of straggler nodes on the convergence rate of the proposed csI-ADMM on synthetic dataset.}
    \label{fig_straggler}
 \end{figure}
It is clear to see that the incremental algorithms including sI-ADMM and WADMM, are more communication effective than the gossip-based benchmarks, such as DADMM, DGD and EXTRA. And the proposed sI-ADMM leads to communication saving for both Hamiltonian based and non-Hamiltonian based token traversing test networks. Meanwhile, it keeps the same test error level compared with benchmarks. This is because that in each iteration, only one communication channel is being used for the incremental methods. Besides, with a fixed token traversing pattern, the proposed sI-ADMM is more balanced in visiting frequency of agents when compared with W-ADMM.

Further, we evaluate robustness vs. the straggler nodes in terms of running time in Fig. \ref{fig_performance_usps} (e). 
Here, running time is defined as the experimental time including both communication time among agents and response time for updating all variables.
Meanwhile, we consider the existence of $N$ stragglers and $S_i=1$ ECN straggler attached to agent $i$ in the test networks. As expected, the baseline uncoded scheme (i.e., sI-ADMM) has worse accuracy performance as delay increases. The proposed csI-ADMM with Cyclic or Fractional schemes has faster response in the presence of stragglers and are not influenced by the delay of the stragglers.
In larger test networks, Fig. \ref{fig_performance_ijcnn1} presents some experimental results based on dataset ijcnn1 and the same performance can be observed as well.

In addition, to investigate the convergence speed vs. straggler nodes trade-off for the proposed csI-ADMM, the impact of number of straggler nodes on the convergence speed is shown in Fig. \ref{fig_straggler}. 
To achieve good performance, we perform 10 independent experiment runs with the same simulation setup on synthetic data and take average for presentation. 
We can see that, with an increasing number of straggler nodes, the convergence speed decreases. This is because increasing the number of straggler nodes decreases the allowable mini-batch size allocated in each iteration and therefore affects the convergence speed.
This is consistent with the analysis in sub-Section \ref{sub-sec:impact} that there exists one trade-off between tolerated straggling number and convergence speed. For the proposed MDS-based csI-ADMM, if we pursue robustness against more straggler nodes, its convergence speed degrades.

\section{Conclusion} \label{conclusion}

We have studied decentralized consensus optimization with ADMM in edge computing enabled large-scale networks. 
An error-control-code based stochastic incremental ADMM algorithm has been proposed to reduce the communication cost for exchanging intermediate model variables with tolerance to link failures and straggler nodes.
We have theoretically analyzed the convergence and communication properties of the proposed csI-ADMM, which show that csI-ADMM reaches a $O(\frac{1}{\sqrt{k}})$ rate and  $O(\frac{1}{\upsilon ^2}) $ communication cost, respectively. 
Moreover, the relation between convergence speed and the number of straggler nodes has also been presented.
Simulation experiments have shown that the proposed csI-ADMM algorithm is more effective in reducing both response time and communication cost while retaining the test loss level, compared with benchmark approaches.

\bibliography{reflibb}

\begin{thebibliography}{10}
\providecommand{\url}[1]{#1}
\csname url@samestyle\endcsname
\providecommand{\newblock}{\relax}
\providecommand{\bibinfo}[2]{#2}
\providecommand{\BIBentrySTDinterwordspacing}{\spaceskip=0pt\relax}
\providecommand{\BIBentryALTinterwordstretchfactor}{4}
\providecommand{\BIBentryALTinterwordspacing}{\spaceskip=\fontdimen2\font plus
\BIBentryALTinterwordstretchfactor\fontdimen3\font minus
  \fontdimen4\font\relax}
\providecommand{\BIBforeignlanguage}[2]{{%
\expandafter\ifx\csname l@#1\endcsname\relax
\typeout{** WARNING: IEEEtran.bst: No hyphenation pattern has been}%
\typeout{** loaded for the language `#1'. Using the pattern for}%
\typeout{** the default language instead.}%
\else
\language=\csname l@#1\endcsname
\fi
#2}}
\providecommand{\BIBdecl}{\relax}
\BIBdecl

\bibitem{iot_device}
\BIBentryALTinterwordspacing
``Internet of things-number of connected devices worldwide 2015-2025,'' 2016.
  [Online]. Available:
  \url{https://www.statista.com/statistics/471264/iot-number-of-connected-devices-worldwide/}
\BIBentrySTDinterwordspacing

\bibitem{big_data2}
S.~Deng, H.~Zhao, W.~Fang, J.~Yin, S.~Dustdar, and A.~Y. Zomaya, ``Edge
  intelligence: the confluence of edge computing and artificial intelligence,''
  \emph{IEEE Internet Things J.}, 2020.

\bibitem{wadmm}
X.~{Mao}, K.~{Yuan}, Y.~{Hu}, Y.~{Gu}, A.~H. {Sayed}, and W.~{Yin}, ``Walkman:
  A communication-efficient random-walk algorithm for decentralized
  optimization,'' \emph{IEEE Trans. Signal Process.}, vol.~68, pp. 2513--2528,
  2020.

\bibitem{pwadmm}
Y.~Ye, H.~Chen, Z.~Ma, and M.~Xiao, ``Decentralized consensus optimization
  based on parallel random walk,'' \emph{IEEE Commun. Lett.}, 2019.

\bibitem{wpg}
X.~Mao, Y.~Gu, and W.~Yin, ``Walk proximal gradient: An energy-efficient
  algorithm for consensus optimization,'' \emph{IEEE Internet Things J.},
  vol.~6, no.~2, pp. 2048--2060, 2018.

\bibitem{DGD}
K.~Yuan, Q.~Ling, and W.~Yin, ``On the convergence of decentralized gradient
  descent,'' \emph{SIAM J. Optim.}, vol.~26, no.~3, pp. 1835--1854, 2016.

\bibitem{EXTRA}
W.~Shi, Q.~Ling, G.~Wu, and W.~Yin, ``{EXTRA}: An exact first-order algorithm
  for decentralized consensus optimization,'' \emph{SIAM J. Optim.}, vol.~25,
  no.~2, pp. 944--966, 2015.

\bibitem{COCA}
W.~{Li}, Y.~{Liu}, Z.~{Tian}, and Q.~{Ling}, ``Communication-censored
  linearized {ADMM} for decentralized consensus optimization,'' \emph{IEEE
  Trans. Signal Inf. Process. Netw.}, vol.~6, pp. 18--34, 2020.

\bibitem{DADMM}
W.~{Shi}, Q.~{Ling}, K.~{Yuan}, G.~{Wu}, and W.~{Yin}, ``On the linear
  convergence of the {ADMM} in decentralized consensus optimization,''
  \emph{IEEE Trans. Signal Process.}, vol.~62, no.~7, pp. 1750--1761, April
  2014.

\bibitem{smart_grid}
W.~{Ma}, J.~{Wang}, V.~{Gupta}, and C.~{Chen}, ``Distributed energy management
  for networked microgrids using online {ADMM} with regret,'' \emph{IEEE Trans.
  Smart Grid}, vol.~9, no.~2, pp. 847--856, 2018.

\bibitem{wsn}
I.~D. {Schizas}, A.~{Ribeiro}, and G.~B. {Giannakis}, ``Consensus in ad hoc
  {WSNs} with noisy links—part {I}: Distributed estimation of deterministic
  signals,'' \emph{IEEE Trans. Signal Process.}, vol.~56, no.~1, pp. 350--364,
  2008.

\bibitem{jacobi_admm}
W.~Deng, M.-J. Lai, Z.~Peng, and W.~Yin, ``Parallel multi-block {ADMM} with
  $o(1 / k)$ convergence,'' \emph{J. Sci. Comput.}, vol.~71, no.~2, pp.
  712--736, May 2017.

\bibitem{feder_learning}
J.~Kone{\v{c}}n{\`y}, H.~B. McMahan, F.~X. Yu, P.~Richt{\'a}rik, A.~T. Suresh,
  and D.~Bacon, ``Federated learning: Strategies for improving communication
  efficiency,'' \emph{arXiv preprint arXiv:1610.05492}, 2016.

\bibitem{d-admm}
J.~F.~C. {Mota}, J.~M.~F. {Xavier}, P.~M.~Q. {Aguiar}, and M.~{Puschel},
  ``{D-ADMM}: A communication-efficient distributed algorithm for separable
  optimization,'' \emph{IEEE Trans. Signal Process.}, vol.~61, no.~10, pp.
  2718--2723, May 2013.

\bibitem{gadmm}
A.~{Elgabli}, J.~{Park}, A.~S. {Bedi}, M.~{Bennis}, and V.~{Aggarwal},
  ``Communication efficient framework for decentralized machine learning,'' in
  \emph{Proc. 2020 54th Annual Conference on Information Sciences and Systems
  (CISS)}, 2020, pp. 1--5.

\bibitem{cola}
W.~{Li}, Y.~{Liu}, Z.~{Tian}, and Q.~{Ling}, ``{COLA}: Communication-censored
  linearized {ADMM} for decentralized consensus optimization,'' in \emph{Proc.
  2019 IEEE International Conference on Acoustics, Speech and Signal Processing
  (ICASSP)}, 2019, pp. 5237--5241.

\bibitem{qsgd}
D.~Alistarh, D.~Grubic, J.~Li, R.~Tomioka, and M.~Vojnovic, ``{QSGD}:
  Communication-efficient {SGD} via gradient quantization and encoding,'' in
  \emph{Proc. Adv. Neural Inf. Process. Syst.}, 2017, pp. 1709--1720.

\bibitem{quantized_admm}
S.~{Zhu} and B.~{Chen}, ``Quantized consensus by the {ADMM}: Probabilistic
  versus deterministic quantizers,'' \emph{IEEE Trans. Signal Process.},
  vol.~64, no.~7, pp. 1700--1713, 2016.

\bibitem{sparsified_sgd}
S.~U. Stich, J.-B. Cordonnier, and M.~Jaggi, ``Sparsified {SGD} with memory,''
  in \emph{Proc. Adv. Neural Inf. Process. Syst.}, 2018, pp. 4447--4458.

\bibitem{qsparse}
D.~Basu, D.~Data, C.~Karakus, and S.~Diggavi, ``{Qsparse-local-SGD}:
  Distributed sgd with quantization, sparsification and local computations,''
  in \emph{Proc. Adv. Neural Inf. Process. Syst.}, 2019, pp. 14\,695--14\,706.

\bibitem{compressed_commu}
A.~Koloskova, S.~U. Stich, and M.~Jaggi, ``Decentralized stochastic
  optimization and gossip algorithms with compressed communication,''
  \emph{arXiv preprint arXiv:1902.00340}, 2019.

\bibitem{speed-up}
K.~{Lee}, M.~{Lam}, R.~{Pedarsani}, D.~{Papailiopoulos}, and K.~{Ramchandran},
  ``Speeding up distributed machine learning using codes,'' \emph{IEEE Trans.
  Inf. Theory}, vol.~64, no.~3, pp. 1514--1529, 2018.

\bibitem{gradient_coding}
R.~Tandon, Q.~Lei, A.~G. Dimakis, and N.~Karampatziakis, ``Gradient coding:
  Avoiding stragglers in distributed learning,'' in \emph{Proc. International
  Conference on Machine Learning}, 2017, pp. 3368--3376.

\bibitem{jingyue}
J.~{Yue} and M.~{Xiao}, ``Coded decentralized learning with gradient descent
  for big data analytics,'' \emph{IEEE Commun. Lett.}, vol.~24, no.~2, pp.
  362--366, 2020.

\bibitem{rscode}
W.~Halbawi, N.~Azizan, F.~Salehi, and B.~Hassibi, ``Improving distributed
  gradient descent using reed-solomon codes,'' in \emph{Proc. 2018 IEEE Int.
  Symp. Inf. Theory (ISIT)}, 2018, pp. 2027--2031.

\bibitem{cyclic_mds}
N.~Raviv, R.~Tandon, A.~Dimakis, and I.~Tamo, ``Gradient coding from cyclic
  {MDS} codes and expander graphs,'' in \emph{Proc. International Conference on
  Machine Learning}, 2018, pp. 4305--4313.

\bibitem{sgc_straggler}
R.~Bitar, M.~Wootters, and S.~El~Rouayheb, ``Stochastic gradient coding for
  straggler mitigation in distributed learning,'' \emph{IEEE Journal on
  Selected Areas in Information Theory}, 2020.

\bibitem{wang2019erasurehead}
H.~Wang, Z.~Charles, and D.~Papailiopoulos, ``Erasurehead: Distributed gradient
  descent without delays using approximate gradient coding,'' \emph{arXiv
  preprint arXiv:1901.09671}, 2019.

\bibitem{ldgm_code}
S.~Horii, T.~Yoshida, M.~Kobayashi, and T.~Matsushima, ``Distributed stochastic
  gradient descent using {LDGM} codes,'' in \emph{Proc. 2019 IEEE Int. Symp.
  Inf. Theory (ISIT)}, 2019, pp. 1417--1421.

\bibitem{ouyang_admm}
H.~Ouyang, N.~He, L.~Tran, and A.~Gray, ``Stochastic alternating direction
  method of multipliers,'' in \emph{Proc. International Conference on Machine
  Learning}, 2013, pp. 80--88.

\bibitem{lian2018asynchronous}
X.~Lian, W.~Zhang, C.~Zhang, and J.~Liu, ``Asynchronous decentralized parallel
  stochastic gradient descent,'' in \emph{Proc. International Conference on
  Machine Learning}, 2018, pp. 3043--3052.

\bibitem{amiri2019computation}
M.~M. Amiri and D.~G{\"u}nd{\"u}z, ``Computation scheduling for distributed
  machine learning with straggling workers,'' \emph{IEEE Trans. Signal
  Process.}, vol.~67, no.~24, pp. 6270--6284, 2019.

\bibitem{ferdinand2020anytime}
N.~Ferdinand, H.~Al-Lawati, S.~C. Draper, and M.~Nokleby, ``Anytime minibatch:
  Exploiting stragglers in online distributed optimization,'' \emph{arXiv
  preprint arXiv:2006.05752}, 2020.

\bibitem{ye_isit}
Y.~Ye, H.~Chen, M.~Xiao, M.~Skoglund, and H.~V. Poor, ``Privacy-preserving
  incremental {ADMM} for decentralized consensus optimization,'' \emph{arXiv
  preprint arXiv:2003.10615}, 2020.

\bibitem{sp-cycle}
J.~N. Al-Karaki and A.~E. Kamal, ``Routing techniques in wireless sensor
  networks: a survey,'' \emph{IEEE Wirel. Commun.}, vol.~11, no.~6, pp. 6--28,
  2004.

\bibitem{usps}
Z.~Kang, K.~Grauman, and F.~Sha, ``Learning with whom to share in multi-task
  feature learning.'' in \emph{Proc. International Conference on Machine
  Learning}, 2011.

\bibitem{libsvm}
C.~Chang and C.~Lin, ``{LIBSVM}: A library for support vector machines,''
  \emph{ACM Trans. Intell. Syst. Technol.}, vol.~2, no.~3, pp. 1--27, 2011.

\end{thebibliography}
\bibliographystyle{IEEEtran} 


\appendices
\section{Proof of Theorem \ref{theorem1} }
\label{secondAppendix}
The augmented Lagrangian (\ref{eq:lagrangian}) can be rewritten as:
\begin{equation} \label{eq:re_lar}
\mathcal{L}_{\rho}(\bm{x},\bm{y},z)=\sum _{i=1}^{N}f_i(x_i) + \frac{\rho}{2}\norm{\mathbbm{1}\otimes z - \bm{x} + \frac{\bm{y}}{\rho}}^2 - \frac{\norm{\bm{y}}^2}{2\rho}.
\end{equation}
From the optimality condition of (\ref{new_x}), we can derive
 \begin{equation}
     \begin{aligned}\label{grad_1}
        &\mathcal{G}_{i_k}(x_{i_k}^{k};\bm \xi_{i_k}^k ) - y_{i_k}^k =\rho\left(z^k-x_{i_k}^{k+1}\right) - \tau^k \left(x_{i_k}^{k+1} - x_{i_k}^k\right) \\
        &= \frac{1}{\gamma^k} \left(y_{i_k}^{k+1} - y_{i_k}^k\right) - \tau^k \left(x_{i_k}^{k+1} - x_{i_k}^k \right).
    \end{aligned} 
 \end{equation}
Taking expectation over both sides in (\ref{grad_1}), we can get
\begin{equation} \label{eq:stocha_gradient}
    \begin{aligned}
        &\mathbbm{E}\left[\mathcal{G}_{i_k}(x_{i_k}^{k};\bm \xi_{i_k}^k ) - y_{i_k}^k\right]=\nabla f_{i_k}(x_{i_k}^{k}) - y_{i_k}^k \\
        &\qquad \qquad = \frac{1}{\gamma^k} \left(y_{i_k}^{k+1} - y_{i_k}^k\right) - \tau^k \left(x_{i_k}^{k+1} - x_{i_k}^k \right).
    \end{aligned}
\end{equation}
Before the proof of Theorem \ref{theorem1}, we first start with the following Lemma \ref{lemma_2}, which is useful for proving the convergence rate of the proposed algorithm when strong convexity is considered. For the simplicity of notation, we use $\mathcal{G}_{i_k}(x_{i_k}^k)$ instead of $\mathcal{G}_{i_k}(x_{i_k}^k;\bm \xi_{i_k}^k )$, and for all parameters, $\epsilon_1, \epsilon_2, \epsilon_3, \text{and } \epsilon_4 > 0$, define
\begin{equation}
    \begin{aligned}
    &D_1 \overset{\Delta}{=} \frac{\left|1-\gamma^k \right|}{\gamma^k};\notag\\
    &D_2 \overset{\Delta}{=} \left|\frac{\tau^k}{N\rho} - \frac{1}{N} + \frac{1}{N^2}\right|;\\
    &D_3 \overset{\Delta}{=} \frac{1}{2\rho \gamma^k}+\frac{1}{2N^2\rho} + \frac{1}{2\rho (\gamma^k) ^2} - \frac{1}{N\rho};\\
    &C_1 \overset{\Delta}{=}D_1 \frac{1}{2\epsilon_1}  -\left(\frac{\mu}{2} -\frac{3\rho}{2}\right);\\
    &C_2 \overset{\Delta}{=} D_1 \frac{\epsilon_1}{2} -D_3 + \frac{1}{2\rho}D_2 + \frac{1}{2N\rho \epsilon_4}; \\
    &C_3 \overset{\Delta}{=} -\left(\frac{\tau^k}{2} + \frac{\rho}{2N^2} +\frac{\tau^k}{N}\right) +  \frac{1}{2N\epsilon_2} + \frac{1}{2N\epsilon_3} + \frac{L^2}{2\rho} + \frac{\rho}{2} D_2;\\
    &C_4 \overset{\Delta}{=} \frac{\epsilon_2}{2N};\notag\\
    &C_5 \overset{\Delta}{=} \frac{\epsilon_3}{2N} + \frac{\epsilon_4}{2N\rho}.
    \end{aligned}
\end{equation}
\begin{lemma}\label{lemma_2}
   For $ k=mN + i$ where $m\in\mathbbm{N}$ and $ i \in\{1,...,N\}$, we have
\begin{align}\label{eq:lemma1}
        &\mathbbm{E} \left[f_{i_k}(x_{i_k}^{k+1} ) - f_{i_k}(x_{i_k} ) +\left \langle y_{i_k}, z^k - x_{i_k}^{k+1}  \right\rangle \right]   \notag\\
        &\leq\frac{\tau^k}{2} \left(\norm{x_{i_k} - x_{i_k}^k}^2 - \norm{x_{i_k} - x_{i_k}^{k+1}}^2\right) \notag\\
        &\quad+ \frac{1}{ 2 \rho \gamma^k}\left(\norm{y_{i_k}-y_{i_k}^{k} }^2 - \norm{y_{i_k} - y_{i_k}^{k+1}}^2\right) \notag\\
        &\quad+ C_1\norm{x_{i_k}^{k+1} - x_{i_k}}^2 + C_2 \norm{y_{i_k}^{k+1} - y_{i_k}^k}^2 \notag\\
        &\quad+ C_3 \norm{x_{i_k}^{k+1} - x_{i_k}^k}^2 +C_4\mathbbm{E}\norm{\nabla f_i(x_{i_k}^k) - \mathcal{G}_{i_k}(x_{i_k}^{k})}^2 \notag\\ 
        &\quad+C_5 \mathbbm{E}\norm{\nabla f_{i_k}(x_{i_k}^k)}^2 + \rho \norm{z- x_{i_k}}^2 .
 \end{align} 
\end{lemma}
\begin{IEEEproof}
  Using strong convexity of $f_i(x_i)$, we can derive 
\begin{align}\label{eq:strong_conv_2}
     &f_{i_k}(x_{i_k}^{k+1}) -f_{i_k}(x_{i_k}) + \left\langle x_{i_k}^{k+1} - x_{i_k} , -y_{i_k}^{k+1}\right\rangle  \notag\\
     &\leq \left\langle x_{i_k}^{k+1} - x_{i_k}, \nabla f_{i_k}(x_{i_k}^{k+1}) - y_{i_k}^{k+1} \right\rangle - \frac{\mu}{2} \norm{x_{i_k}^{k+1} - x_{i_k}}^2\notag\\
    &= \underbrace{\left\langle x_{i_k}^{k+1} - x_{i_k}, \nabla f_{i_k}(x_{i_k}^{k}) - y_{i_k}^{k} \right\rangle}_{\mathcal{A}} \notag\\ 
    &\quad +\left\langle x_{i_k}^{k+1} - x_{i_k}, \nabla f_{i_k}(x_{i_k}^{k+1}) - \nabla f_{i_k}(x_{i_k}^{k})\right \rangle \notag\\
    &\quad + \left\langle x_{i_k}^{k+1} - x_{i_k}, y_{i_k}^{k} - y_{i_k}^{k+1} \right\rangle -\frac{\mu}{2} \norm{x_{i_k}^{k+1} - x_{i_k}}^2. 
    \end{align} 
Taking expectation over term $\mathcal{A}$ in (\ref{eq:strong_conv_2}) gives
    \begin{align} \label{e_t1}
    &\mathbbm{E} \left [\mathcal{A} \right ]  \notag\\
    &=\mathbbm{E}\left[\left\langle x_{i_k}^{k+1} - x_{i_k}, \nabla f_{i_k}(x_{i_k}^{k}) - \mathcal{G}_{i_k}(x_{i_k}^{k}) + \mathcal{G}_{i_k}(x_{i_k}^{k}) - y_{i_k}^{k} \right\rangle \right] \notag\\
    &\overset{(a)}{=} \tau^k \left\langle x_{i_k}^{k+1} - x_{i_k},  x_{i_k}^{k} - x_{i_k}^{k+1}  \right \rangle\notag\\ &~~~~+
     \frac{1}{\gamma^k} \left\langle x_{i_k}^{k+1} - x_{i_k},  y_{i_k}^{k+1} - y_{i_k}^{k}  \right \rangle \notag \\
    &\overset{(b)}{=}\frac{\tau^k}{2}\left(\norm{x_{i_k}-x_{i_k}^{k}}^2 - \norm{x_{i_k} - x_{i_k}^{k+1}}^2 - \norm{x_{i_k}^{k} - x_{i_k}^{k+1}}^2\right)   \notag\\
    &~~~~+\frac{1}{\gamma^k} \left \langle x_{i_k}^{k+1} - x_{i_k}, y_{i_k}^{k+1} - y_{i_k}^{k}  \right\rangle,
    \end{align}
where (a) holds because of (\ref{eq:stocha_gradient}) and (b) uses the cosine identity
$\norm{b + c}^2 - \norm{a+c}^2 = \norm{b-a}^2 + 2\left\langle a+c, b-a \right\rangle$\label{(b)}.
From (\ref{new_y}), we derive: 
    \begin{align}\label{e_t2}
    &\left\langle y_{i_k}^{k+1} - y_{i_k},-z^k + x_{i_k}^{k+1} \right \rangle 
    =  \frac{1}{\rho \gamma^k} \left \langle y_{i_k}^{k+1} - y_{i_k}, y_{i_k}^{k} - y_{i_k}^{k+1}  \right\rangle \notag\\
    &= \frac{1}{2\rho \gamma^k}\left(\norm{y_{i_k}-y_{i_k}^{k}}^2 - \norm{y_{i_k} - y_{i_k}^{k+1}}^2 - \norm{y_{i_k}^{k} - y_{i_k}^{k+1}}^2\right),
    \end{align} 
and
    \begin{align}\label{e_t3}
     \left\langle z^k - z,y_{i_k}^{k+1} \right\rangle  = &\left\langle z^{k+1} - z,y_{i_k}^{k+1}\right \rangle + \left\langle z^{k} - z^{k+1},y_{i_k}^{k+1}  \right\rangle  \notag\\
     \overset{(c)}{\leq }&\left\langle z^{k} - z^{k+1},y_{i_k}^{k+1}  \right\rangle - \frac{\rho}{2}\norm{z^{k+1} - x_{i_k}^{k+1}}^2 \notag \\&+\frac{\rho}{2}\norm{z - x_{i_k}^{k+1}}^2,
    \end{align} 
where (c) is due to the optimality of $z^{k+1}$ for update (\ref{old_z}), i.e., $ \langle z^{k+1} - x_{i_k}^{k+1},y_{i_k}^{k+1}  \rangle + \frac{\rho}{2}\big\| z^{k+1} - x_{i_k}^{k+1}\big\| ^2 \leq \langle z -x_{i_k}^{k+1} ,y_{i_k}^{k+1}  \rangle + \frac{\rho}{2}\big\| z - x_{i_k}^{k+1}\big\| ^2$.
Moreover,
\begin{align}
   \frac{\rho}{2}\norm{z - x_{i_k}^{k+1}}^2 &= \frac{\rho}{2}\norm{z - x_{i_k} + x_{i_k} - x_{i_k}^{k+1}}^2\notag\\
   &\overset{(d)}{\leq} \rho \norm{z - x_{i_k}}^2 + \rho \norm{x_{i_k}^{k+1} - x_{i_k} }^2,
\end{align}
where (d) is due to $(a+b)^2\leq 2a^2 + 2b^2$.
Thus, taking the summation of the inequalities (\ref{eq:strong_conv_2}), (\ref{e_t2}) and (\ref{e_t3}), we obtain (\ref{T1_3}).
\begin{figure*}[t] 
    \begin{align}\label{T1_3}
    &f_{i_k}(x_{i_k}^{k+1})  -f_{i_k}(x_{i_k}) + \left\langle y_{i_k}, z^k - x_{i_k}^{k+1}  \right \rangle \leq \frac{\tau^k}{2} \left(\norm{x_{i_k} - x_{i_k}^k}^2 - \norm{x_{i_k} - x_{i_k}^{k+1}}^2\right)  + \frac{1}{ 2 \rho \gamma^k}\left(\norm{y_{i_k}-y_{i_k}^{k} }^2  - \norm{y_{i_k} - y_{i_k}^{k+1}}^2\right)\notag\\
     &\quad + \frac{1-\gamma^k}{\gamma^k} \left \langle x_{i_k}^{k+1} - x_{i_k}, y_{i_k}^{k+1} - y_{i_k}^{k} \right \rangle  - \frac{\mu - 2\rho}{2} \norm{x_{i_k}^{k+1} - x_{i_k}}^2  + \rho \norm{z - x_{i_k}}^2  +  \left\langle x_{i_k}^{k+1} - x_{i_k}, \nabla f_{i_k}(x_{i_k}^{k+1}) - \nabla f_{i_k}(x_{i_k}^{k}) \right \rangle   \notag\\
     & \quad- \frac{\tau^k}{2}\norm{x_{i_k}^k - x_{i_k}^{k+1}}^2 -\frac{1}{2\rho \gamma^k} \norm{y_{i_k}^{k} - y_{i_k}^{k+1}}^2  -  \underbrace{\frac{\rho}{2}\norm{z^{k+1} - x_{i_k}^{k+1}}^2}_{\mathcal{B}}  + \underbrace{\left\langle z^{k} - z^{k+1},y_{i_k}^{k+1}   \right\rangle}_{\mathcal{C}},
    \end{align} 
     \hrulefill
    \end{figure*}
For term $\mathcal{B}$, we have
    \begin{align}
    \mathcal{B}
    &=\frac{\rho}{2} \norm{z^{k+1} - z^{k}}^2 +\frac{\rho}{2} \norm{z^{k} - x_{i_k}^{k+1}}^2 \notag\\
    &\quad + \rho    \left \langle z^{k+1} - z^{k}, z^{k} - x_{i_k}^{k+1}   \right \rangle \notag\\
    &=  \frac{\rho}{2} \norm{z^{k+1} - z^{k}}^2 + \frac{1}{2\rho (\gamma^k) ^2} \norm{y_{i_k}^k -    y_{i_k}^{k+1}}^2  \notag\\
    &\quad + \frac{1}{\gamma^k}\left \langle z^{k+1} - z^{k}, y_{i_k}^{k+1} - y_{i_k}^{k}\right\rangle \notag\\
    &= \left(\frac{1}{2N^2\rho} + \frac{1}{2\rho  (\gamma^k) ^2}- \frac{1}{N\rho \gamma^k}\right)\norm{y_{i_k}^{k+1} - y_{i_k}^{k}}^2\notag \\
    &\quad -\left(\frac{1}{N^2}- \frac{1}{\gamma^k N}\right)\left\langle x_{i_k}^{k+1} - x_{i_k}^{k}, y_{i_k}^{k+1} - y_{i_k}^{k}  \right \rangle\notag\\ 
    &\quad +\frac{\rho}{2N^2}\norm{x_{i_k}^{k+1} - x_{i_k}^{k}}^2.
    \end{align} 
And for term $\mathcal{C}$, it can be written as (\ref{eqc}).
\begin{figure*}[t]
    \begin{align}\label{eqc}
    \mathcal{C}
    &=\left\langle z^{k} - z^{k+1},y_{i_k}^{k}  \right \rangle - \left \langle z^{k} - z^{k+1},y_{i_k}^{k} - y_{i_k}^{k+1}  \right \rangle  = -\frac{1}{N}\left\langle   x_{i_k}^{k+1} - x_{i_k}^{k} ,\mathcal{G}_{i_k}(x_{i_k}^{k})-\frac{1}{\gamma^k}\left(y_{i_k}^{k+1} - y_{i_k}^{k}\right)  +\tau^k \left(x_{i_k}^{k+1}- x_{i_k}^{k}\right) \right\rangle \notag\\ &~~~+\frac{1}{N\rho} \left\langle y_{i_k}^{k+1} - y_{i_k}^{k},\mathcal{G}_{i_k}(x_{i_k}^{k})  -\frac{1}{\gamma^k}(y_{i_k}^{k+1}- y_{i_k}^{k})+\tau^k(x_{i_k}^{k+1} - x_{i_k}^{k}) \right \rangle   +\frac{1}{N} \left\langle x_{i_k}^{k+1} - x_{i_k}^{k}y_{i_k}^{k} - y_{i_k}^{k+1} \right\rangle + \frac{1}{N\rho} \norm{y_{i_k}^{k+1}- y_{i_k}^{k}}^2 \notag\\
    &= -\frac{1}{N}  \left\langle x_{i_k}^{k+1} - x_{i_k}^{k}, \mathcal{G}_{i_k}(x_{i_k}^{k}) \right\rangle + \frac{1}{N\rho} \left\langle y_{i_k}^{k+1} - y_{i_k}^{k}, \mathcal{G}_{i_k}(x_{i_k}^{k}) \right \rangle  \left(\frac{1}{N\gamma^k} + \frac{\tau^k}{N\rho} - \frac{1}{N}\right)  \left\langle x_{i_k}^{k+1} - x_{i_k}^{k}, y_{i_k}^{k+1} - y_{i_k}^{k} \right\rangle\notag\\
    &~~~ -\frac{\tau^k}{N}\norm{x_{i_k}^{k+1} - x_{i_k}^{k}}^2 - \left(\frac{1}{N\rho \gamma^k} - \frac{1}{N\rho}\right)\norm{y_{i_k}^{k+1} - y_{i_k}^{k}}^2.
    \end{align}
    \hrulefill
    \end{figure*} 
Then equation (\ref{T1_3}) can be transferred to \eqref{e_t123}.
\begin{figure*}[t]
    \begin{align}\label{e_t123}
    &f_{i_k}(x_{i_k}^{k+1}) - f_{i_k}(x_{i_k}) + \left\langle y_{i_k}, z^k - x_{i_k}^{k+1} \right \rangle\leq   \frac{\tau^k}{2} \left(\norm{x_{i_k} - x_{i_k}^k}^2 - \norm{x_{i_k} - x_{i_k}^{k+1}}^2\right) + \frac{1}{ 2 \rho \gamma^k}\left(\norm{y_{i_k}-y_{i_k}^{k} }^2 - \norm{y_{i_k} - y_{i_k}^{k+1}}^2\right)\notag\\
    &\quad+\underbrace{\frac{1-\gamma^k}{\gamma^k}  \left\langle x_{i_k}^{k+1} - x_{i_k}, y_{i_k}^{k+1} - y_{i_k}^{k}  \right\rangle}_{\mathcal{D}}  +  \underbrace{\left\langle x_{i_k}^{k+1} - x_{i_k}, \nabla f_{i_k}(x_{i_k}^{k+1}) - \nabla f_{i_k}(x_{i_k}^{k})  \right\rangle}_{\mathcal{E}} -\left(\frac{\mu}{2}-\rho \right)\norm{x_{i_k}^{k+1} - x_{i_k}}^2 \notag\\
    &\quad- \left(\frac{\tau^k}{2} + \frac{\rho}{2N^2} + \frac{\tau^k}{N}\right) \norm{x_{i_k}^{k+1} - x_{i_k}^{k}}^2
    -\left(\frac{1}{2\rho \gamma^k} + \frac{1}{2N^2 \rho} + \frac{1}{2\rho  (\gamma^k) ^2}  - \frac{1}{N\rho}\right)\norm{y_{i_k}^{k+1} - y_{i_k}^{k}}^2  + \rho \norm{z - x_{i_k}}^2 \notag\\
    &\quad+\underbrace{\left(\frac{\tau^k}{N\rho} -\frac{1}{N} + \frac{1}{N^2}\right)  \left\langle x_{i_k}^{k+1} - x_{i_k}^{k}, y_{i_k}^{k+1} - y_{i_k}^{k}  \right\rangle}_{\mathcal{F}} -\underbrace{ \frac{1}{N}  \left\langle x_{i_k}^{k+1} - x_{i_k}^{k}, \mathcal{G}_{i_k}(x_{i_k}^{k})   \right\rangle}_{\mathcal{G}}+ \underbrace{\frac{1}{N\rho}  \left\langle y_{i_k}^{k+1} - y_{i_k}^{k},  \mathcal{G}_{i_k}(x_{i_k}^{k}) \right \rangle}_{\mathcal{H}}. 
    \end{align} 
    \hrulefill
 \end{figure*}
Here, using Young's inequality, the components $\mathcal{D}$--$\mathcal{F}$ of (\ref{e_t123}) can be upper bounded as follows:
    \begin{align}\label{eqD}
    &\mathcal{D}\leq
    \frac{\left|1-\gamma^k\right|}{\gamma^k} \left(\frac{1}{2\epsilon_1}\norm{x_{i_k}^{k+1} - x_{i_k}}^2 +\frac{\epsilon_1}{2}\norm{y_{i_k}^{k+1} - y_{i_k}^{k}}^2\right),  
\end{align}
and
    \begin{align}\label{eqE}
    \mathcal{E}&\leq \frac{\rho}{2} \norm{x_{i_k}^{k+1} - x_{i_k}}^2 + \frac{1}{2\rho}\norm{ \nabla f_{i_k}(x_{i_k}^{k+1}) - \nabla f_{i_k}(x_{i_k}^{k})}^2 \notag \\
    &\leq \frac{\rho}{2} \norm{x_{i_k}^{k+1} - x_{i_k}}^2 + \frac{ L^2}{2\rho}\norm{ x_{i_k}^{k+1} - x_{i_k}^{k}}^2, 
    \end{align}
 and
    \begin{align}\label{eqF}
    \mathcal{F}\leq & \left|\frac{\tau^k}{N\rho} - \frac{1}{N} + \frac{1}{N^2}\right| \notag \\ & \qquad\times\left(\frac{\rho}{2}\norm{x_{i_k}^{k+1} - x_{i_k}^{k}}^2  
     + \frac{1}{2\rho}\norm{y_{i_k}^{k+1} - y_{i_k}^{k}}^2\right).
    \end{align}
    Taking expectation over $\mathcal{G}_{i_k}(x_{i_k}^{k})$ with respect to the mini-batch $\bm{\xi}_{i_k}^k$ in term $\mathcal{G}$ and $\mathcal{H}$, we can derive
    \begin{align}\label{eqG}
    \mathbbm{E}& \left [ \mathcal{G}\right ] =
    \mathbbm{E}\left[-\frac{1}{N} \left \langle x_{i_k}^{k+1} - x_{i_k}^{k}, \mathcal{G}_{i_k}(x_{i_k}^{k})-\nabla f_{i_k}(x_{i_k}^{k}) \right \rangle \right] \notag\\
    &\quad- \frac{1}{N} \left\langle x_{i_k}^{k+1} - x_{i_k}^{k}, \nabla f_{i_k}(x_{i_k}^{k})\right \rangle \notag\\
    &\leq \left(\frac{1}{2N\epsilon_2} + \frac{1}{2N\epsilon_3}\right) \norm{x_{i_k}^{k+1} - x_{i_k}^{k}}^2   + \frac{\epsilon_3}{2N} \norm{\nabla f_{i_k}(x_{i_k}^k)}^2\notag\\
    &\quad +\frac{\epsilon_2}{2N}\mathbbm{E}\left[\norm{\nabla f_{i_k}(x_{i_k}^{k})-\mathcal{G}_{i_k}(x_{i_k}^{k})}^2 \right], 
    \end{align}
 and
    \begin{align}\label{eqH}
    &\mathbbm{E}  \left[ \mathcal{H}\right ] 
        \leq\frac{1}{2N\rho \epsilon_4} \norm{y_{i_k}^{k+1} - y_{i_k}^{k}}^2 + \frac{\epsilon_4}{2N\rho}  \norm{\nabla f_{i_k}(x_{i_k}^k)}^2.
    \end{align} 
Thus, by plugging (\ref{eqD})-(\ref{eqH}) into (\ref{e_t123}) and taking the expectation, we can obtain the desired result (\ref{eq:lemma1}).
This completes the proof.
\end{IEEEproof}


Then based on the result given in Lemma 1, we present the proof for Theorem 2. 
We first give the following assumptions:
  \begin{align}
    & \epsilon_1 = c_1\sqrt{k},~
   \epsilon_2 = \frac{1}{c_2 \sqrt{k}},~
    \epsilon_3 = \frac{1}{c_3\sqrt{k}},~\epsilon_4 = \frac{1}{c_4 \sqrt{k}},\notag\\
    &
    \tau^k = c_{\tau}\sqrt{k},~
   \gamma^k = \frac{c_{\gamma}}{\sqrt{k}}, ~3\rho < \mu,\notag\\
    &\frac{1}{\mu-3\rho}< c_1  c_{\gamma } < \frac{1}{\rho},~ c_{\tau}\left(N+1\right) >c_2+c_3,
    \end{align}
where $c_1, ~c_2, ~c_3, ~c_4, ~c_{\tau}, \text{ and }c_{\gamma}  >0, \text{ are constants }, k=mN + i$ is the $k $-th iteration number, cycle index $m\in\{0, ...,T-1\}$ and $i\in\{1, ...,N\}$.
When $k$ is sufficiently large, we have $C_1 <0, C_2<0, \text{ and } C_3<0$.

Based on Lemma \ref{lemma_2}, we have 
    \begin{align}\label{eq:single_exp}
    &\mathbbm{E}\left[f_{i_k}(x_{i_k}^{k+1}) - f_{i_k}(x_{i_k}) +  \left\langle y_{i_k}, z^k - x_{i_k}^{k+1} \right \rangle\right]  \notag\\
    &~~~~~~~~~~~\leq\frac{\tau^k}{2} \left(\norm{x_{i_k} - x_{i_k}^k}^2 - \norm{x_{i_k} - x_{i_k}^{k+1}}^2\right) \notag\\
    &~~~~~~~~~~~\quad+ \left(\frac{\epsilon_3}{2N} + \frac{\epsilon_4}{2N\rho}\right)\norm{\nabla f_{i_k}(x_{i_k}^k)}^2+ \rho \norm{z- x_{i_k}}^2 \notag\\
    &~~~~~~~~~~~\quad+\frac{1}{ 2 \rho \gamma^k}\left(\norm{y_{i_k}-y_{i_k}^{k} }^2 - \norm{y_{i_k} - y_{i_k}^{k+1}}^2\right)\notag\\
    &~~~~~~~~~~~\quad+ \frac{\epsilon_2}{2N}\mathbbm{E}\left[\norm{\nabla f_{i_k}(x_{i_k}^{k})-\mathcal{G}_{i_k}(x_{i_k}^{k})}^2 \right]. 
    \end{align} 
Through summing (\ref{eq:single_exp}) from $k=1$ to $k=TN, T \in \mathbbm{N}$, 
we obtain
\begin{align}\label{eq:all_cyc}
    &\frac{1}{TN} \sum_{m=0}^{T-1} \sum_{i=1}^{N} {\mathbbm{E}\left[{f_i(x_i^{k+1}) - f_i(x_i)} +  \left\langle  y_i, z^k - x_i^{k+1} \right\rangle\right]  }
     \notag\\
    &\leq\frac{1}{TN}  \sum_{m=0}^{T-1} \sum_{i=1}^{N}\left[\frac{\tau^k}{2}\left (\norm{x_{i} - x_{i}^k}^2 - \norm{x_{i} - x_{i}^{k+1}}^2 \right) \right. \notag\\
    &\qquad\quad\qquad\quad + \frac{1}{ 2 \rho \gamma^k}\left(\norm{y_{i}-y_{i}^{k} }^2 - \norm{y_{i} - y_{i}^{k+1}}^2\right)  \notag\\
    &\qquad\quad\qquad\quad \left. +\left(\frac{\epsilon_3}{2N} + \frac{\epsilon_4}{2N\rho}\right)\norm{\nabla f_{i}(x_{i}^k)}^2+ \rho \norm{z- x_{i_k}}^2 \right. \notag\\
    &\qquad\quad\qquad\quad \left. + \frac{\epsilon_2}{2N}\mathbbm{E}\left[\norm{\nabla f_{i}(x_{i}^{k})-\mathcal{G}_{i}(x_{i}^{k})}^2\right]\right].
\end{align}     
For equation (\ref{eq:all_cyc}) at the optimal solution $(x_{i}, z) = (x_{i}^*, z^*),\text{ where } z^*=x_i^*, i \in \mathcal{N} $, we can derive that $\forall y_{i}\in \mathbbm{R}^{p\times d}$, 
 the above inequality is true,
 hence it also holds in the ball $\mathcal{B}_{\beta}= \{y_{i}: \norm{y_{i}} \leq \beta  \}$.
 Combining with the fact that the optimal solution must also be feasible, it follows that
 \begin{align} \label{eq:sup}
     &\mathop{ \text{sup}}\limits_{\beta \in \mathcal{B}_{\beta}} \left\{ \frac{1}{TN}  \sum_{m=0}^{T-1} \sum_{i=1}^{N} \mathbbm{E}\left[{f_{i}(x_{i}^{k+1}) - f_{i}(x_{i}^*) +  \left\langle  y_{i}, z^k - x_{i}^{k+1} \right\rangle}\right]  \right\}\notag\\
     & =\mathop{ \text{sup}}\limits_{\beta \in \mathcal{B}_{\beta}} \left\{ \frac{1}{TN} \sum_{m=0}^{T-1} \sum_{i=1}^{N} \mathbbm{E}\left[{f_{i}(x_{i}^{k+1})}\right] - \frac{1}{N}\sum_{i=1}^{N} f_{i}(x_{i}^*)  \right.  \notag\\
     &\qquad\qquad \left. + \left \langle  y_{i},  \frac{1}{TN} \sum_{m=0}^{T-1} \sum_{i=1}^{N} \mathbbm{E}\left[z^k - x_{i}^{k+1}\right]\right \rangle \right \} \notag \\
     & = \frac{1}{TN} \sum_{m=0}^{T-1} \sum_{i=1}^{N} {\mathbbm{E}\left[ f_{i}(x_{i}^{k+1})\right]}  - \frac{1}{N} \sum_{i=1}^{N} f_{i}(x_{i}^*) \notag\\
     &\qquad\qquad +\beta \mathbbm{E}\left[ \norm{ \frac{1}{TN}  \sum_{m=0}^{T-1} \sum_{i=1}^{N} (z^k - x_{i}^{k+1}) }\right].
 \end{align}
Thus, using (\ref{eq:sup}), we obtain 
\begin{align}\label{eq:con_rate_1}
    & \frac{1}{TN}  \sum_{m=0}^{T-1} \sum_{i=1}^{N} { \mathbbm{E} \left [f_{i}(x_{i}^{k+1}) - f_{i}(x_{i}^*)\right ] }   \notag\\
    &~~~~~~~~~\qquad+\beta \mathbbm{E} \left[\norm{ \frac{1}{TN}  \sum_{m=0}^{T-1} \sum_{i=1}^{N} (z^k - x_{i}^{k+1}) }\right]\notag\\
    &\leq \frac{1}{TN}  \sum_{m=0}^{T-1} \sum_{i=1}^{N}\left[\frac{c_{\tau}\sqrt{k}}{2} \left( \norm{x_{i}^* - x_{i}^k}^2 - \norm{x_{i}^* - x_{i}^{k+1}}^2 \right)  \right. \notag\\
    &~~~~~~~~~ \qquad\left.+\frac{\sqrt{k}}{ 2 \rho c_{\gamma}}\left(\norm{y_{i}-y_{i}^{k} }^2 - \norm{y_{i} - y_{i}^{k+1}}^2\right)\right.\notag\\
    &~~~~~~~~~ \qquad\left.+\left(\frac{1}{2N c_3\sqrt{k}} + \frac{1}{2N\rho c_4 \sqrt{k}}\right)\phi + \frac{\delta ^2 }{2MN c_2 \sqrt{k}}\right]  \notag\\
    &\leq \frac{1}{TN} \left(\frac{c_{\tau}ND_{\mathcal{X}}^2\sqrt{TN} }{2} + \frac{2N\sqrt{TN}}{  \rho c_{\gamma}} \beta ^2 \right) \notag\\
    &\quad+\frac{1}{TN} \sum_{k=1}^{TN} \left[\left(\frac{1}{2N c_3\sqrt{k}} + \frac{1}{2N\rho c_4\sqrt{k}}\right)\phi+\frac{\delta ^2 }{2MN c_2 \sqrt{k}}\right]\notag\\
    &\leq \frac{c_{\tau} ND_{\mathcal{X}}^2 }{2\sqrt{TN}}  + \frac{2N\beta^2}{\rho c_{\gamma} \sqrt{TN}}  +\left(\frac{1}{N c_3} + \frac{1}{N\rho c_4}\right)\frac{\phi}{\sqrt{TN}} \notag\\
    &\quad+\frac{\delta ^2 }{M N c_2 \sqrt{TN}}.
\end{align} 
Especially, applying $c_1 = 1,~c_2 = \frac{1}{{N}},~c_3 = \frac{1}{{N}},~c_4=\frac{1}{\rho {N}},~ \frac{1}{\mu-3\rho}< c_{\gamma } < \frac{1}{\rho}, \text{ and } c_{\tau} > \frac{2}{\left(N+1\right){N}}$ in (\ref{eq:con_rate_1}), the final result in (\ref{eq:theorem}) can be derived.
This completes the proof.


\end{document}